\documentclass[12pt]{amsart}
\usepackage{amsthm,amsmath,amsxtra,amscd,amssymb}
\usepackage{tikz}
\usepackage{latexsym,amsmath,amssymb,mathrsfs}
\usepackage{amsfonts,eucal,amsthm,graphicx,color}

\numberwithin{equation}{section}

\newcommand{\mt}{{\sf m}}
\newcommand{\nt}{{\sf n}}
\newcommand{\se}{{\sf u}}
\newcommand{\ba}{{\bf a}}
\newcommand{\bb}{{\bf b}}
\newcommand{\bc}{{\bf c}}
\newcommand{\tht}{\mathbf{\theta}}

\newcommand{\ve}{{\varepsilon}}

\newcommand{\calC}{{\mathcal C}}

\newcommand{\calG}{{\mathcal G}}

\def\beq{\begin{equation}}
\def\eeq{\end{equation}}

\newcommand\Res{\operatorname{Res}}

\theoremstyle{plain}
\newtheorem{thm}{Theorem}[section]
\newtheorem{lm}[thm]{Lemma}

\theoremstyle{definition}

\newtheorem{rem}[thm]{Remark}

\newenvironment{aside}{\begin{quote}\sffamily}{\end{quote}}

\newcommand {\BC}   {\mathbb C}

\newcommand {\BH}   {\mathbb H}

\newcommand {\BR}   {\mathbb R}
\newcommand {\BP}   {\mathbb P}

\newcommand {\BT}   {\mathbb T}
\newcommand {\bA}   {\mathbf{A}}
\newcommand {\bU}   {\mathbf{U}}
\newcommand {\bZ}   {\mathbf{Z}}

\newcommand {\bm}{ \mathbf{m}}
\newcommand {\bn}{ \mathbf{n}}
\newcommand {\bu}{ \mathbf{u}}
\newcommand {\bl}{ \mathbf{l}}

\newcommand {\bz}{ \mathbf{z}}

\newcommand {\bS}{ \mathbf{S}}
\newcommand {\BS}   {\mathbb S}

\newcommand {\BZ}   {\mathbb Z}

\newcommand {\CalC} {\mathcal C}

\newcommand {\CalE} {\mathcal E}
\newcommand {\CalF} {\mathcal F}
\newcommand {\CalG} {\mathcal G}
\newcommand {\CalH} {\mathcal H}

\newcommand {\CalJ} {\mathcal J}

\newcommand {\CalL} {\mathcal L}
\newcommand {\CalM} {\mathcal M}
\newcommand {\CalN} {\mathcal N}
\newcommand {\CalO} {\mathcal O}

\newcommand {\CalS} {\mathcal S}

\newcommand {\CalU} {\mathcal U}

\newcommand {\CalW} {\mathcal W}
\newcommand {\CalZ} {\mathcal Z}

\newcommand {\zb} {{\bar z}}

\newcommand {\be}{{\sf{e}}}

\newcommand{\fo}{\vert\kern -.03in\_}

\newcommand {\ii} {\mathrm{i}}
\newcommand{\Tr}{\text{Tr}}
\newcommand{\pa}{{\partial}}
\newcommand{\G}{\Gamma}
\newcommand {\g}{\gamma}
\newcommand{\s}{\sigma}
\begin{document}
\title[Lefschetz thimbles in Sigma models, I]{Towards Lefschetz thimbles in Sigma models, I}
\author{Igor Krichever}

\address{Department of Mathematics, 
Columbia University, New York NY 10027, USA\\
Center for Advanced Studies, Skoltech, Moscow 143026 Russia\\
Kharkevich Institute for Information Transmission Problems, Moscow 127051 Russia\\
{\tt E-mail: krichev@math.columbia.edu}}

\author{Nikita Nekrasov}

\address{Simons Center for Geometry and Physics, 
Stony Brook University, Stony Brook NY 11794-3636, USA
\\
Center for Advanced Studies, Skoltech, Moscow 143026 Russia\\
Kharkevich Institute for Information Transmission Problems, Moscow 127051 Russia\\
{\tt E-mail: nnekrasov@scgp.stonybrook.edu}}

\maketitle

We study two dimensional path integral Lefschetz thimbles, i.e. the possible path integration contours. 
Specifically,
in the examples of the $O(N)$ and ${\BC\BP}^{N-1}$  models, we find a large class of complex critical points of 
the sigma model actions which are relevant for the theory in finite volume at finite temperature, with various chemical potentials corresponding to the symmetries of the models. In this paper we discuss the case of the $O(2m)$ and the ${\BC\BP}^{N-1}$ models in the sector of zero instanton charge, as well as some solutions of the $O(2m+1)$ model. 
The  ${\BC\BP}^{N-1}$-model for all instanton charges and a more general class of solutions of the $O(N)$-model with odd $N$ will be discussed in the 
forthcoming paper.

\section{Introduction}

The partition function of a quantum field theory is formally given by the path integral
\beq
Z_{\hbar} ({\bf t}) = \int_{{\CalF}}\, [D{\Phi}] \, e^{-\frac{S_{\bf t}({\Phi})}{\hbar}}
\label{eq:pf1}
\eeq
over some space $\CalF$ of fields.  The correlation functions are defined in a similar manner:
\beq
\langle {\CalO}_{1}(x_{1}) \ldots {\CalO}_{r}(x_{r}) \rangle_{\bf t} = \frac{1}{Z_{\hbar} ({\bf t})} \int_{{\CalF}}\, [D{\Phi}] \, e^{-\frac{S_{\bf t}({\Phi})}{\hbar}} \, {\CalO}_{1}(x_{1}) \ldots {\CalO}_{s}(x_{s})
\label{eq:corrfun}
\eeq
 In order to study the analytic properties of these quantities, as the parameters $t$ are varied, a useful trick is to deform the contour of integration, 
 so as if it were a middle dimensional cycle in the complexification ${\CalF}^{\BC}$ of the space of fields (which has a meaning in the A model \cite{Witten:2010zr}, in the B model \cite{Hori:2000ck}, and in the $\Omega$-deformed gauge theories \cite{Yagi:2014toa, Nekrasov:2018pqq}). 
 There may be many cycles ${\gamma}_{\ba}$
 for which the integral \eqref{eq:pf1} converges. For finite dimensional integrals,
 these are classified by the relative homology group
 $H_{\rm middle} ( {\CalF}^{\BC} , {\CalF}^{\BC}_{-} ; {\BZ})$ where
 ${\CalF}^{\BC}_{-}$ is the set of ${\Phi} \in {\CalF}^{\BC}$ for which
 ${\rm Re}(S_{t}({\Phi})/{\hbar}) \gg 0$. For generic $t$ there is a basis $({\gamma}_{\ba})$ in $H_{\rm middle} ( {\CalF}^{\BC} , {\CalF}^{\BC}_{-} ; {\BZ})$ of the so-called Lefschetz thimbles. The thimble
 $\gamma_{\ba}$ corresponds to the critical point ${\ba}$ of
 $S_{t}({\Phi})$, and is the union of gradient trajectories of ${\rm Re} \left(S_{t}({\Phi})/{\hbar}\right)$ emanating
 from ${\ba}$. The original path integral decomposes as the sum
 \beq
 Z_{\hbar}(t) = \sum_{\ba} n_{\ba} I_{\hbar}^{\ba}(t)
 \eeq
 where
 \beq
 I_{\hbar}^{\ba}(t) = \int_{{\gamma}_{\ba}} [D{\Phi}] \, e^{-\frac{S_{t}({\Phi})}{\hbar}}
\label{eq:pf2}
\eeq
is the integral over the Lefschetz thimble corresponding to the critical
point, \emph{the complex classical solution} ${\ba}$. The multiplicities $n_{\ba}$
are integers. For small variations of $t$ the multiplicities are constant, however, as $t$ cross certain hypersurfaces (the walls of marginal stability) the integers $n_{\ba}$ may jump. This is known as the Stokes phenomenon.

In this paper we are studying the critical points ${\ba}$ in several examples of field theories. 

\subsubsection{Fields and symmetries}

The space of fields ${\CalF}$ in each model is the set of maps
of some source manifold $\Sigma$ to the target space $X$.
We will be avoiding the rather toxic questions about the precise degree of smoothness of the maps
relevant for the path integral measure.  Of course, one should look into the cutoff version of the path integral measure, where the naive microscopic action $S_{t}({\Phi})$ is replaced
by some  $S_{t({\wedge})}({\Phi}_{\wedge})$ where the characteristic momenta of the fields ${\Phi}_{\wedge}$ are less than $\wedge$, and the couplings $t({\wedge})$ are made depend on $\wedge$ in such a way, that the limit ${\wedge} \to \infty$ makes the correlation
functions \eqref{eq:corrfun} finite, for macroscopic separations between the points $x_{1}, \ldots , x_{s}$. Hopefully, for the problem of classification of the
possible path integral contours for the asymptotically free theories, such as the sigma models on the positively curved target spaces, the more rigorous treatment will give a similar result. 

To probe the symmetries of the theory one also studies the integrals over the spaces ${\rm Maps}_{h}({\Sigma}, X)$ of twisted maps. Here
$h: {\pi}_{1}({\Sigma}) \to H$  denotes a homomorphism of the fundamental group of $\Sigma$ to the group of symmetries of $X$ (and additional structures on $X$). The $h$-twisted maps are
the ${\pi}_{1}({\Sigma})$-equivariant maps  $f: {\tilde\Sigma} \to X$
of the universal cover of $\Sigma$, obeying:
\beq
f( {\gamma} \cdot p) = h({\gamma}) \cdot f(p)
\label{eq:twistedmap}
\eeq
In other words, if a point $\xi \in \Sigma$ and a representative $f({\xi}) \in X$ is chosen, then the twisted map $f$ in the neighborhood $U$ of $\xi$ is a well-defined map $U \to X$, however its continuation to the whole of $\Sigma$ is multi-valued, up to the action of
$h ({\pi}_{1}({\Sigma})) \subset G$ on $X$.

Another possibility presented by the symmetries of $X$ are the defect operators, classified by all the homotopy groups ${\pi}_k (H)$. For example, 
the elements of ${\pi}_{{\rm dim}({\Sigma})-1}(H)$ classify local twist operators, these can be described as the instruction to perform the path
integral over the space ${\Gamma}({\Sigma}, X \times_{H} {\bf H})$ of sections of the fiber bundle associated with the principal $H$-bundle $\bf H$
over $\Sigma$, corresponding to the element 
\beq
c \in H^{{\rm dim}{\Sigma}}\left({\Sigma}, {\pi}_{{\rm dim}({\Sigma})-1}(H)\right)\ .
\label{eq:gensw}
\eeq

\subsubsection{Quantum mechanics}
The case of $\Sigma = {\bS}^{1}$ was discussed in \cite{Nekrasov:2018pqq}, where the class of quantum mechanical (= $\Sigma$ is one-dimensional) models was considered. The target $(X, {\omega})$ is a symplectic manifold such that its complexification $(X^{\BC}, {\omega}^{\BC})$ is an algebraic integrable system
${\pi}: X^{\BC} \to U^{\BC} \approx {\BC}^{r}$ with the fibers $J_{u} = {\pi}^{-1}(u)$ being Lagrangian polarized abelian variety for generic $u \in U^{\BC}$. The action $S_{t}({\Phi})$ is given by:
\beq
S_{\bf t}({\Phi}) = \oint_{\Sigma} d^{-1} {\omega}^{\BC} - \sum_{k=1}^{r} t_{k}
\oint u_{k}\left( s \right) ds
\eeq
where $s \sim s+1$ is a parameter on $\Sigma$, $u_{k}$ are the global coordinates on $U$ -- the Hamiltonians of the integrable system, $\bf t$ is the set of \emph{times}, or generalized inverse temperatures, so that the path integral \eqref{eq:pf1} represents the trace of the complexified evolution operator:
\beq
Z_{\hbar} ({\bf t}) = {\Tr}_{\CalH} \, e^{- \frac{1}{\hbar}\sum_{k} t_{k} {\hat H}_{k} }
\label{eq:gentr}
\eeq

{}Let $\Xi \subset U$ denote the discriminant, i.e. the set of singular fibers $J_{u}$.  Pick a basepoint $u_{*} \in U\backslash \Xi$. Let
${\Gamma}$ be the monodromy group, i.e. the image of the based fundamental group
${\pi}_{1}   \left( U^{\BC} \backslash {\Xi}, u_{*} \right)$ in the group $ Sp(2r, {\BZ}) \times {\BZ}^{f}$ of affine transformations of the fiber
$H_{1}(J_{u_{*}}, {\BZ})$, preserving the symplectic structure given by the intersection form (with the help of the polarization).

The critical points ${\ba}$ in these examples are
classified by the $\Gamma$-orbits $[c]$ of the homology classes $c \in H_{1}\left( {\pi}^{-1}(u_{*}), {\BZ} \right)$
under the action of the monodromy group.  Let ${\Gamma}_{[c]} \subset \Gamma$ denote the stabilizer of $c$, and ${\tilde U}_{[c]} = {\tilde U}/{\Gamma}_{[c]}$ be the associated quotient of the universal cover. Clearly, ${\tilde U}_{n[c]} = {\tilde U}_{[c]}$, for any integer $n \neq 0$. 
Let $P := PH_{1}(J_{u_{*}}, {\BZ})$ denotes the set of primitive homology classes (i.e. the homology classes which are not  multiples of the others). Let ${\CalU}_{\rho} = {\tilde U}_{{\rho}}$, ${\rho} \in P$.  For $n \in {\BZ}$, ${\bf t} \in {\BC}^{r}$ define the \emph{superpotential}, the holomorphic function on  ${\CalU}_{\rho}$, given by
\beq
{\CalW}_{n, {\bf t}} = n \oint_{\rho}  d^{-1}{\omega}^{\BC} - \sum_{k=1}^{r} t_{k} u_{k}\ .
\label{eq:superpot}
\eeq
The set  
\beq
{\CalC} = \bigsqcup\limits_{{\rho} \in P} \ {\CalC}_{\rho}\ , \qquad {\CalC}_{\rho} \subset {\CalU}_{\rho}
\eeq
of Lefschetz thimbles of quantized algebraic integrable system can be viewed as a discrete subset of 
\beq
{\CalU} = \bigsqcup\limits_{{\rho} \in P} \ {\CalU}_{\rho}
\eeq
For $\rho \neq 0$ the set ${\CalC}_{\rho}$ is the set of critical points of \emph{superpotential} \eqref{eq:superpot} on ${\CalU}_{\rho}$. 
There is a subtlety at $\rho =0$, where ${\CalU}_{0} \approx U$. The superpotential is this case becomes simply a linear combination of the regular functions $u_k$, which has no critical points. However, on physical grounds, we should include in ${\CalC}$ the set ${\Xi}_{\rm max} \subset {\Xi}$ of maximally degenerate
fibers. These correspond to the degenerate orbits of the Hamiltonian vector field generated by $\sum_k t_k u_k$ viewed as a function on the whole phase space $X^{\BC}$. 

There is a simple modification of the problem in case of the systems with symmetries, preserving the Hamiltonians $h_k$. Instead of the space of loops
${\CalF} = LX$ and its complexification ${\CalF}^{\BC} = LX^{\BC}$ one considers, as in \eqref{eq:twistedmap}, the spaces 
${\CalF}_{[h]} = \{ \, x(s) \, | \, 0 \leq s \leq 1\, ,\ x(1) = h \cdot x(0) \}$ of twisted loops and
its complexification ${\CalF}_{[h_{c}]}^{\BC}$. Here $[h]$ is a conjugacy class in the group $H$ of symplectic symmetries of $X$, $[h_{c}]$
is a conjugacy class in the complex group $H^{\BC}$ of symplectic holomorphic symmetries of $X^{\BC}$. Since the group preserves the Hamiltonians, it
acts on the fibers $J_u$. In case the group $H$ is a Lie group
generated by some Hamiltonians, the twisted case is equivalent to the untwisted one, up to a redefinition of the times $\bf t$. The
case of the discrete group $H$ is quite interesting, and is not completely covered in the literature. It can again be reduced
to \eqref{eq:superpot} with $P$ being the space of equivalence classes of $h$-twisted loops on $J_u$. 

\subsubsection{Organization of the paper}

We are studying the two dimensional field theories, with the target spaces an odd dimensional sphere ${\bS}^{2m-1}$, 
or its quotient with respect to the $U(1)$-action producing the complex projective space ${\BC\BP}^{n-1}$. 
The complexifications of these spaces contain, as real sections, other interesting symmetric manifolds, 
including the Lobachevsky space, the anti-de Sitter, and de Sitter spaces.

In section \ref{sec:sigma models} we introduce the Lagrangians of the sigma models in two dimensions, and their realization through the (linear) sigma models with constraints and gauge symmetries. 
We then discuss the models with twisted boundary conditions and their complexification. We then present the first hints the structure similar to the Eqs. \eqref{eq:superpot} can be expected in the case of the two dimensional sigma models: in the restricted class of the \emph{folded string solutions}, we recover a special class of algebraic integrable systems: Gaudin model, i.e. Hitchin system at genus zero  with punctures, as in \cite{N95}, albeit  with irregular singularities. 

The section \ref{sec:fermi} introduces the main tool in our analysis: the Fermi-curves. First, we remind the construction of the Fermi-curve for a small perturbation of a constant Schr{\"o}dinger potential. We show that the double points are resolved by the Fourier modes of the potential $u(z, {\zb})$.

The section \ref{sec:alg-int} reverse-engineers the Schr{\"o}dinger potential $u(z, {\zb})$
from the Fermi-curve of finite genus. We characterize the latter as an analytic curve with an additional structure: an involution with two fixed points and a set of meromorphic differentials $\Omega$, $\Omega^{\pm}$ with specified properties.

The section \ref{sec:selfc} relates  Schr{\"o}dinger operator $- {\Delta} + u$ to its solutions $\psi$, ${\Delta}{\psi} = u {\psi}$, 
as specified by the sigma model equations of motion. We show that this relation implies the additional structure on the Fermi-curve: a meromorphic function $E$. Moreover, we find the \emph{superpotential} ${\CalW}$ whose critical points correspond to the double-periodic solutions 
of the sigma model equations of motion, much like the superpotential \eqref{eq:superpot} we find in the quantum mechanical model. In fact, we make the relation between the two explicit.

The section \ref{sec:conc} presents our conclusions and directions of future research. 
In particular, we comment on the finite-dimensional approximations of the field theory configuration space, inspired by the finite-gap solutions.

\section{Sigma models}\label{sec:sigma models}

Let $X$, the \emph{target space} be a Riemannian manifold with the metric $g = g_{mn}(X) dX^{m} dX^{n}$. The action $S_{t}({\Phi})$ of the sigma model will be taken to be
\beq
S_{t}({\Phi}) = \int_{\Sigma} \ \sqrt{h} h^{\alpha\beta} \ g_{mn} {\pa}_{\alpha} X^{m} {\pa}_{\beta} X^{n}
\label{eq:acsigma}
\eeq
where $\left( X^{m} (z, {\zb}) \right) $  parametrizes the map ${\Phi}: {\Sigma} \to X$, while $t$ stands for other parameters described below.  
The Lagrangian of the two dimensional sigma model depends on the conformal class of metric $ds^{2}_{\Sigma} = h_{ab}d{\xi}^{a} d{\xi}^{b}$, $h_{ab} \sim e^{2{\psi}} h_{ab}$.

Let $\Sigma$ denote a two-dimensional torus ${\bS}^{1} \times {\bS}^{1}$. 
Let $x,y$ denote the real coordinates on $\Sigma$, with the periods $1$, i.e. $x \sim x+m$, $y \sim y + n$ with $m,n \in {\BZ}$. 

on $\Sigma$, which is 
parametrized by the complex number $\tau = {\tau}_{1} + {\ii} {\tau}_{2}$, with ${\tau}_{2} >0$, via
\beq
ds^{2}_{\Sigma} \propto (dx + {\tau} dy ) (dx + {\bar\tau}y) = dz d{\zb}
\label{eq:fltmtr}
\eeq
where  
 $z =  x + {\tau} y$, ${\zb} = x + {\bar\tau} y$
denote the holomorphic and anti-holomorphic coordinates on $\Sigma$, respectively.

In what follows we often use the notation ${\omega}_{x} = 1$, 
${\omega}_{y} = {\tau}$ to denote the periods, and ${\bar\omega}_{x} = 1, {\bar\omega}_{y} = {\bar\tau}$ to denote the conjugate periods.

$t$ stands for the parameters of the conformal structure 
$\sqrt{h}h^{\alpha\beta}$ of $\Sigma$, the parameters of the metric $g$, as well as the twist parameters. 
The latter occur when the metric $g$ has isometries. Let  $H$ denote the group of symmetries of $g$. We deform the theory
by turning on a flat $H$-connection $A$
\beq
S_{t}({\Phi}; A) = \int_{\Sigma} \ dzd{\bar z} \ g_{mn} ({\pa}_{z} X^{m} + A^{\ba}_{z} V_{\ba}^{m} )( {\pa}_{\zb} X^{n} + A^{\ba}_{\zb} V^{n}_{\ba} )
\label{eq:acsigmag}
\eeq
where $A^{\ba}dz+  {\bar A}^{\ba}d{\zb}$ is the $H$-connection form,
with $\ba = 1, \ldots , {\rm dim}H$, and $V_{\ba} \in {\rm Vect}(X)$ the generators
of $H$ acting by isometries of $X$. The invariance of the path integral measure under the local transformations of the map ${\Phi}: {\Sigma} \to X$ by the isometry group $H$ implies that the  correlation functions depend only on the gauge equivalence class of $A \sim h^{-1}A h + h^{-1} dh$.

{}We can rewrite \eqref{eq:acsigmag} using the background-independent form of $\Sigma$, i.e. in the coordinates $x,y$:
\begin{multline}
S_{t}({\Phi}; A) = \frac{1}{{\tau}_{2}} \, \int_{{\BR}^{2}/{\BZ}^{2}}\, dxdy \, {\CalL} \, , \\
\, \\
 {\CalL} \, = \, g_{mn}(X)\, \left( {\tau} {\nabla}_{x} X^{m} - {\nabla}_{y} X^{m} \right) \left( {\bar\tau} {\nabla}_{x} X^{n} -  {\nabla}_{y} X^{n} \right)
\label{eq:sigmac}
\end{multline}
  with 
  \beq
  {\nabla}_{\alpha}X^{m} = {\pa}_{\alpha}X^{m} + A_{\alpha}^{\ba}V_{\ba}^{m}(X)\, , \ \alpha = x,y \ .
  \label{eq:covder}
  \eeq
  
When the $H$-connection $A$ is flat,
\beq
dA^{\ba} + \frac 12 f^{\ba}_{{\bb}{\bc}} A^{\bb} \wedge A^{\bc} = 0\ , 
\eeq
the partition function \eqref{eq:pf1} can be cast in the Hamiltonian form:
\beq
  {\CalZ} ( A; {\tau}, {\bar\tau} ) = {\Tr}_{{\CalH}_{g_{x}-{\rm twisted}}} \, \left( \, g_{y} \, q^{H_{+}} {\bar q}^{H_{-}}  \right)
  \label{eq:pf3}
\eeq
where $q = e^{2\pi\ii \tau}, {\bar q} = e^{-2\pi\ii{\bar\tau}}$, 
\begin{multline}
H_{\pm} = \frac{1}{4\pi} \left( {\hat H} \pm {\hat P} \right) \, , \\
 \ g_{x} = P\exp \int_{0}^{1} dx A_{x}\, , \ g_{y} = P\exp \int_{0}^{1} dy A_{y}\, ,
\end{multline}
are the light-cone Hamiltonians, and the $H$-twists, respectively, and ${\CalH}_{g_{x}-{\rm twisted}}$ denotes the space of states of the theory obtained by quantizing the space of $g_{x}$-twisted loops into the target space $X$. 

An interesting aspect of theories with the symmetry groups $H$ having nontrivial fundamental group ${\pi}_{1}(H)$ is the possibility of having topologically nontrivial backgrounds, while maintaining the flatness of the background connection $A$.  They 
are in one-to-one correspondence with the elements $c \in H_{2}({\Sigma}, {\pi}_{1}(H))$, cf. \eqref{eq:gensw}, known as the generalized Stiefel-Whitney classes,  for finite
${\pi}_{1}(H)$. For simple Lie group $H$, ${\pi}_{1}(H)$ is identified with the subgroup of the center $Z({\tilde H})$ of its simply-connected cover ${\tilde H}$.  The topologically nontrivial background is defined by studying the path integral over the space ${\CalF}_{c}$ of sections of the $X$-bundle over $\Sigma$, associated to the principal $H$-bundle $P$ over $\Sigma$. The latter can be trivialized over a complement ${\Sigma}\backslash U_p$ to a small neighborhood $U_p$ of a point $p \in {\Sigma}$, as well as over $U_p$ itself. The class $c\in {\pi}_{1}(H)$ is then represented by the loop in $H$ given by the map ${\partial}U_{p} \to H$ comparing the two trivializations $P \vert_{\Sigma \backslash U_p} \times H \times {\Sigma \backslash U_p}$ and $P \vert_{U_{p}} \approx H \times U_{p}$. 

Thus, on the one hand, the path integral over ${\CalF}_{c}$ can be interpreted as $1$-point function on the torus of a local disorder operator ${\CalO}_{c}$, 
\beq
 {\CalZ}_{c} ( A; {\tau}, {\bar\tau} ) = {\Tr}_{{\CalH}_{g_{x}-{\rm twisted}}} \, \left( \,
 {\CalO}_{c}\,  g_{y} \, q^{H_{+}} {\bar q}^{H_{-}} \, \right)
\label{eq:twtr}
\eeq
On the other hand the bundle $P_{c}$ lifts to a trivial bundle ${\tilde\Sigma} \times {\tilde H}$, 
for some isogeny ${\tilde\Sigma} \to {\Sigma}$. 
The flat $H$-connection on $\Sigma$ can be viewed as the flat $\tilde H$-connection on $\tilde \Sigma$, 
equivariant with respect to the ${\pi}_{1}(H)$-action. 

Classically, we only see $H$ as the symmetry of the theory, as it is the symmetry of the target space $X$. Quantum mechanically, however, the group $H$ may act on the Hilbert space of the theory projectively, i.e. ${\CalH}$ is really a ${\tilde H}$-representation. 
Then \eqref{eq:twtr} is an important tool for tracking the extension of the symmetry. 

For example, one can reliably demonstrate the existence of BPS solitons in ${\CalN}=2$ supersymmetric ${\BC\BP}^{N-1}$ sigma model  (see, e.g. \cite{Hanany:1997vm}), which transform in the fundamental representations ${\wedge}^{l}{\BC}^{N}$ of the group ${\tilde H} = SU(N)$. Thus the Hilbert space of
the theory contains excitations transforming under the extended symmetry.

 \subsection{Complexification}
  
The partition function \eqref{eq:pf3} admits an analytic continuation in the parameters
$q, {\bar q}, g_{y}$. It would correspond to taking the same path integral over the (twisted) maps
${\Phi}: {\Sigma} \to X$ but now in the presence of deformation of the
action \eqref{eq:sigmac}, where the parameters $\tau$ and
  $\bar\tau$ are not complex conjugates, and the component $A_{y}$  of the background gauge field is complex. Modular invariance then suggests one should be able to promote $A_x$ to the complex gauge field as well. 
  The saddle points of \eqref{eq:sigmac} would, naturally, have $X^m$ complex as well, thus
  prompting the search for the complex solutions of the sigma model equations of motion.
  
  We shall still call ${\omega}_{\alpha}$ and ${\bar\omega}_{\alpha}$, for ${\alpha}=x,y$, the conjugate periods. The conjugation in question corresponds to the symmetry $(x,y) \mapsto (x, -y)$ of the physical torus, not the (artificial in the present context) complex conjugation. 

In what follows we discuss the geometric aspects of the complexification of the fields ${\Phi}$. They are naturally the maps of $\Sigma$ to the complexification $X_{\BC}$ of the original target space. Next we discuss the analytic continuation of the Lagrangians ${\CalL}$ of our sigma models, twisted boundary conditions, and equations of motion. We conclude by the detailed analysis of an interesting reduction of the equations of motion of the $O(N)$ and ${\BC\BP}^{N-1}$ models, the so-called \emph{winding ansatz}. The significance of this ansatz is its algebraic integrability. We find the winding string solutions of the $O(N)$ and ${\BC\BP}^{N-1}$ models are described by an irregular version of the genus zero Hitchin system, a spin zero ${\mathfrak{gl}}_{2}$ Gaudin model. 

\subsubsection{Complexifications of spheres and projective spaces}

The familiar triple ${\BR\BP}^{m}, {\BC\BP}^{m}, {\BH\BP}^{m}$ of spaces with the $O(m), U(m), Sp(m)$ symmetries has an interesting complex version.

For a vector space $L$ over a field $k$, let us denote by $L^{\vee}$ the dual vector space. For $l \in L$, $p \in L^{\vee}$ we denote by $p \cdot l \in k$ the value of $p$ at $l$.

Let $V \approx {\BC}^{m+1}$ be a complex Euclidean space, i.e. a complex vector space with the non-degenerate symmetric form $g \left( {\cdot} ,{\cdot} \right)$. Let $W \approx {\BC}^{n+1}$
be a complex vector space. Finally, let $U \approx {\BC}^{2(n+1)}$ be a complex symplectic vector space, i.e. a complex vector space with the non-degenerate anti-symmetric form ${\omega}( {\cdot}, {\cdot})$. 

Of course, non-canonically, $U = W \oplus W^{\vee}$ for any Lagrangian subspace $W\subset U$, and, in the case of even $n+1$, $V = W \oplus W^{\vee}$ for any maximally isotropic $W \subset V$. 

The complexification $S(V)$ of the sphere $S^{m}$ is the space of vectors $x \in V$, obeying $g(x, x ) = 1$. 

The complexification ${\BP}{\BR}(V)$ of ${\BR\BP}^{m}$ is the quotient of the space of vectors $x \in V$, obeying $g( x,x ) =1$, modulo the ${\BZ}_{2}$-symmetry $x \mapsto -x$. 

The complexification ${\BP}{\BC}(W)$ of ${\BC\BP}^{m}$ is the space of pairs $({\psi}, {\psi}^{\sigma})$, ${\psi} \in W, {\psi}^{\sigma} \in W^{\vee}$, obeying 
\beq
{\psi}^{\sigma} \cdot {\psi} = 1\ , 
\label{eq:momcp}
\eeq
modulo the ${\BC}^{\times}$-action
\beq
({\psi}, \, {\psi}^{\sigma}) \mapsto (t {\psi}, \, t^{-1}{\psi}^{\sigma}),  \qquad t \in {\BC}^{\times}\ . 
\eeq
In other words, ${\BP}{\BC}(W)$ is the holomorphic symplectic quotient $T^{*}W//{\BC}^{\times}$, \eqref{eq:momcp} being the moment map equation. 

The complexification ${\BP}{\BH}(U)$ of ${\BH\BP}^{m}$ is the quotient of the space of pairs $(u_{1}, u_{2})$ with $u_{1,2} \in U$, obeying
\beq
{\omega}(u_{1}, u_{2}) = 1
\label{eq:momhp}
\eeq
by the $SL(2, {\BC})$-action
\beq
(u_{1}, u_{2}) \mapsto (a u_{1} + b u_{2}, cu_{1} + d u_{2} )\ , \qquad ad - bc = 1 \ .
\eeq

\subsubsection{Complexification: the Lagrangians}

The corresponding Lagrangians are written in the terms of the same geometric structures:
\beq
\begin{aligned}
& L_{{\BR\BP}^{m}} = \frac 12 \int_{{\BT}^{2}} \sqrt{h} \left( \, h^{\alpha\beta} g ({\partial}_{\alpha} x, {\partial}_{\beta} x) +  \left( g(x,x) - 1 \right) U\, \right)  \\
& L_{{\BC\BP}^{m}} = \frac 12 \int_{{\BT}^{2}} \sqrt{h} \left( \, h^{\alpha\beta}   D_{\alpha}{\psi}^{\sigma} \cdot D_{\beta} {\psi} + 
 \left( {\psi}^{\sigma} {\cdot} {\psi} - 1 \right)U\,  \right) \\
& \qquad\qquad\qquad\qquad D_{\alpha}{\psi}^{\sigma} =  {\pa}_{\alpha} {\psi}^{\sigma} - {A}_{\alpha} {\psi}^{\sigma} \, , \quad D_{\alpha} {\psi} = {\pa}_{\alpha} {\psi} + {A}_{\alpha} {\psi} \, , \\
& L_{{\BH\BP}^{m}} = \frac{{\ve}^{AB}}{4} \int_{{\BT}^{2}} \sqrt{h} \left( \, h^{\alpha\beta}  {\omega} \left( {\mathfrak{D}}_{\alpha} u_{A} , {\mathfrak{D}}_{\beta} u_{B} \right)   +  \left( {\omega}(u_{A}, u_{B}) - {\ve}_{AB} \right) U\,  \right) \, , \\
& \qquad\qquad\qquad\qquad {\mathfrak{D}}_{\alpha} u_{1} = {\pa}_{\alpha} u_{1}  +  {{A}}_{\alpha} u_{1} + B_{\alpha} u_{2} \, ,  \quad {\mathfrak{D}}_{\alpha} u_{2} = {\pa}_{\alpha} u_{2} + {{C}}_{\alpha} u_{1} - {{A}}_{\alpha} u_{2}  \, ,  \\
\end{aligned}
\label{eq:clac}
\eeq
with ${\ve}^{12} = -{\ve}^{21} = 1$. 
The $S(V)$-model which we shall call the $O(n+1)$-model in what follows has the Lagrangian 
\beq
L_{{\BR\BP}^{n}} = \int_{{\BT}^{2}} \sqrt{h} \left( \frac 12 h^{ab} g ({\partial}_{a} x, {\partial}_{b} x) + \left( g(x,x) - 1 \right) U  \right)
\label{eq:onmodel}
\eeq
which is identical to that of the ${\BR\BP}^{m}$-model. The difference is in gauging.
The ${\BR\BP}^{n}$-model is obtained by identifying the solutions $\left( x(z, {\zb}) \right)$ and $\left( -x(z, {\zb}) \right)$. One is also forced to include the twisted sectors \cite{Orbifolds}, where
$x(z+{\omega}_{\alpha}, {\zb}+{\bar\omega}_{\alpha}) = u_{\alpha} x (z, {\zb})$, with $u_{\alpha} = \pm 1$, $\alpha = 1,2$.

The classical ${\BC\BP}^{m-1}$ model consists of solving the Euler-Lagrange equations following from $L_{{\BC\BP}^{m-1}}$. However, certain care is needed in 
formulating the periodicity conditions. 

The space of fields of the ${\BC\BP}^{m-1}$ model is the space of maps 
\beq
{\CalF} = {\rm Maps}({\Sigma}, {\BC\BP}^{m-1}) = {\rm Maps}\left({\Sigma}, {\bS}^{2m-1}/U(1)\right)\ , 
\eeq
which in turn is the quotient of the space of $U(1)$-equivariant maps 
\beq
{\CalF} = {\rm Maps} \left( P, {\bS}^{2m-1} \right)^{U(1)} / {\CalG}_{P}
\eeq 
of some principal $U(1)$-bundle $P$ over $\Sigma$ to the sphere ${\bS}^{2m-1}$, by the action of the group ${\CalG}_{P} = {\rm Maps}({\Sigma}, U(1))$ of gauge transformations.  

Accordingly, the set of connected components ${\pi}_{0}{\rm Maps}({\Sigma}, {\BC\BP}^{m-1})$ is, for $n\geq 2$, the set of topological classes of $P$, which is isomorphic to ${\BZ}$. In this paper we shall only consider the case of $P = {\Sigma} \times U(1)$, a zero element in $\BZ$. 

In this case we can describe the corresponding complexification as the set of maps ${\psi}: {\Sigma} \to W$, ${\psi}^{\sigma}: {\Sigma} \to W^{\vee}$, 
supplemented with the twisted boundary conditions ${\psi}(z+{\omega}_{\alpha}, {\zb}+{\bar\omega}_{\alpha}) = u_{\alpha}(z, {\bar z}) {\psi} (z, {\zb})$, ${\psi}^{\sigma}(z+{\omega}_{\alpha}, {\zb}+{\bar\omega}_{\alpha}) = u_{\alpha}(z, {\bar z})^{-1} {\psi}^{\sigma} (z, {\zb})$, for $\alpha = 1, 2$ with $u_{\alpha}(z, {\bar z})  \in {\BC}^{\times}$, modulo the identifications $\left( {\psi} , {\psi}^{\sigma} \right) \sim \left( t {\psi}, t^{-1}   {\psi}^{\sigma} \right)$, for any ${\BC}^{\times}$-valued double periodic function $t : {\Sigma} \to {\BC}^{\times}$. The gauge field ${\mathfrak{A}}_{\alpha}$
is required to obey the twisted periodicity condition
\beq
{{A}}_{\alpha}(z+{\omega}_{\beta}, {\bar z}+{\bar\omega}_{\beta}) = {{A}}_{\alpha}(z, {\bar z}) + u_{\beta}(z, {\bar z})^{-1} {\pa}_{\alpha} u_{\beta}(z, {\bar z})\ .
\eeq
We leave the definition of the classical ${\BH\BP}^{n}$ model as an exercise to the reader.

\subsubsection{The principal chiral field models}

More generally we might be interested in the case where $X = {\calG}$ a compact Lie group. In this so-called called principal chiral field model one takes a ${\calG} \times {\calG}$-invariant metric on $X$:
\beq
\label{eq:pcm}
G = {\rm tr} (g^{-1}dg)^{2}\ ,
\eeq
The  principal chiral field model has $H={\calG}\times {\calG}$ group of symmetries.

\subsubsection{Twisted boundary conditions and complexification}

As we explained in the introduction \eqref{eq:acsigmag}, the simplest twisted boundary conditions
correspond to a choice of a flat connection in a principal
$H$-bundle $P_{c}$ over $\Sigma$. 
Practically, for $\Sigma \approx {\bS}^{1} \times {\bS}^{1}$ it means fixing two commuting elements $h_{x}, h_{y} \in H$, $h_{x} h_{y} = h_{y} h_{x}$, up to a simultaneous conjugation $(h_{x}, h_{y}) \equiv ( h^{-1} h_{x} h, h^{-1} h_{y} h)$, $h \in H$.

In the case of $X = {\calG}$ the twisted boundary conditions have the form:
\beq
g(z+1, {\zb}+1) = a_{L} g(z, {\zb})a_{R}^{-1}\, , \ g(z+{\tau} , {\zb} + {\bar\tau})
= b_{L} g(z, {\zb}) b_{R}^{-1}
\eeq
where $a_{L,R}, b_{L,R} \in {\calG}$, and can be simultaneously conjugated
to the maximal torus $T \subset {\calG}$.

In the case of $X = {\bS}^{2m-1}$, the commuting pair of generic twists $h_{x}, h_{y} \in O(V)$, 
determines a decomposition $V\otimes {\BC} = W \oplus W^{\vee}$, such that $h_{x},h_{y}$
can be represented by some commuting unitary operators $a, b \in GL(W)$, $[a,b] = 0$:
\beq
\begin{aligned}
 & h_{x} \left( {\psi} \oplus {\psi}^{\sigma} \right) = a \cdot {\psi} \oplus {\psi}^{\sigma} a^{-1} \, , \\
& h_{y} \left( {\psi} \oplus {\psi}^{\sigma} \right) = b \cdot {\psi} \oplus {\psi}^{\sigma} b^{-1} \, , \\
& \qquad\qquad {\psi} \in W\, , \ {\psi}^{\sigma} \in W^{\vee} \\
& \qquad\qquad g ({\psi}_{1} \oplus {\psi}_{1}^{\sigma} , {\psi}_{2} \oplus {\psi}_{2}^{\sigma}) = 
{\psi}_{2}^{\sigma}\cdot {\psi}_{1} + {\psi}_{1}^{\sigma} \cdot {\psi}_{2} \, . 
\end{aligned}
\label{eq:hxhywws}
\eeq
In the case of $X = {\bS}^{2m}$, the commuting pair of generic twists $h_{x}, h_{y} \in O(2n+1)$, $h_{x}h_{y} = h_{y}h_{x}$, determine a decomposition $V\otimes {\BC} = W \oplus W^{\vee} \oplus {\BC}$,
such that $h_{x},h_{y}$
can be represented by some commuting unitary operators $a, b \in U(W)$, as in \eqref{eq:hxhywws} and act on $\BC$ by multiplication by $\pm 1$. Here the metric on $V$ is expressed as:
\beq
\Vert {\psi}\oplus {\psi}^{\sigma}\oplus {\chi} \Vert^{2}_{g} = 2 {\psi}^{\sigma}\cdot {\psi} + {\chi}^{2}
\eeq
Upon complexification we simply take ${\psi}$, ${\psi}^{\sigma}$, and $\chi$ to be independent $W, W^{\vee}, {\BC}$-valued fields, while $a, b$ become the general commuting elements of $GL(W)$. 
In this paper we shall mostly consider the case, where 
\beq
a = {\rm diag}(a_{1}, \ldots, a_{n}) \, , \ b = {\rm diag}(b_{1}, \ldots, b_{n})\, , 
\label{eq:diagtw}
\eeq
however, the case of Jordan blocks is also quite interesting.

{}In the case of $X = {\BC\BP}^{m-1}$ the boundary conditions
are classified by an element $c = e^{\frac{2\pi \ii p}{m}} \in {\BZ}_{m}$, the second Stiefel-Whitney class of the
$SU(m)/{\BZ}_{m}$ bundle, and a pair $h_{x}, h_{y}$ of the $SU(m)$ matrices obeying
\beq
h_{x}h_{y} = c h_{y}h_{x}
\label{eq:twcomm}
\eeq
up to the simultaneous conjugation $h_{\alpha} \sim h^{-1}h_{\alpha} h$, ${\alpha}=x,y$ with $h \in SU(m)$.
The boundary conditions read, now:
\beq
\begin{aligned}
& {\psi}(z+{\omega}_{\alpha}, {\zb}+{\bar\omega}_{\alpha}) = h_{\alpha} e^{{\ii}{\varphi}_{\alpha}(z,{\zb})} {\psi} (z, {\zb})\, , \\
& {\psi}^{\sigma}(z+{\omega}_{\alpha}, {\zb}+{\bar\omega}_{\alpha}) =  e^{-{\ii}{\varphi}_{\alpha}(z,{\zb})} {\psi}^{\sigma} (z, {\zb}) h_{\alpha}^{-1}\, , \\
& {A}_{a} (z + {\omega}_{\alpha}, {\zb} + {\bar\omega}_{\alpha}) = 
{A}_{a} (z, {\zb}) + {\ii} {\partial}_{a} {\varphi}(z, {\zb}) 
\end{aligned}
\eeq
where $e^{{\ii}{\varphi}_{\alpha}(z,{\zb})}$ is a $U(1)$-valued function. 
Let $l = gcd (p,m)$, and $k = m/l$. Then $h_{1} h_{2}^{k} = h_{2}^{k} h_{1}$, so upon going onto a $k$-fold cover $\tilde\Sigma$ of $\Sigma$ one can obtain the usual twisted boundary conditions. 
Upon complexification the operators $h_{x}, h_{y}$ become the generic $GL(W)$ elements, commuting up to an element $c$ of the center, as in \eqref{eq:twcomm}. 

\subsection{The equations of motion upon complexification}

In the $O(N)$-case we get, for $N$ even
\beq
\begin{aligned}
& {\pa}_{z}{\pa}_{\zb}{\psi} \, = \, U {\psi} \, , \  {\pa}_{z}{\pa}_{\zb}{\psi}^{\sigma} \, = \, U {\psi}^{\sigma}  \, ,  \, \\
& U = - \frac 12 \left( {\pa}_{z} {\psi}^{\sigma} \cdot {\pa}_{\zb} {\psi} + {\pa}_{\zb} {\psi}^{\sigma} \cdot {\pa}_{z} {\psi} \right)\, ,  \\
& \qquad\qquad\qquad\qquad\qquad\qquad {\psi}^{\sigma} \cdot {\psi} = 1\, , 
\end{aligned}
\label{eq:eomeven}
\eeq
for $N$ odd:
\beq
\begin{aligned}
& {\pa}_{z}{\pa}_{\zb}{\psi} \, = \, U {\psi} \, , \  {\pa}_{z}{\pa}_{\zb}{\psi}^{\sigma} \, = \, U {\psi}^{\sigma}  \, ,  \, {\pa}_{z}{\pa}_{\zb}{\chi} =  U {\chi}  \, , \\
& U = - \frac 12 \left( {\pa}_{z} {\psi}^{\sigma} \cdot {\pa}_{\zb} {\psi} + {\pa}_{\zb} {\psi}^{\sigma} \cdot {\pa}_{z} {\psi} \right)  - {\pa}_{z}{\chi} {\pa}_{\zb} {\chi}\, , \\
& \qquad\qquad\qquad\qquad\qquad\qquad {\psi}^{\sigma} \cdot {\psi} + {\chi}^{2} = 1 \ .
\end{aligned}
\label{eq:eomodd}
\eeq
In the case of the ${\BC\BP}^{N-1}$ model, upon complexification, 
the equations of motion read:
\beq
\begin{aligned}
& - \frac 12 \left( D_{\zb} D_{z} + D_{z}D_{\zb} \right) {\psi} + U\, {\psi} = 0\, , \\
& - \frac 12 \left( D_{\zb} D_{z} + D_{z}D_{\zb} \right) {\psi}^{\sigma} + U\, {\psi}^{\sigma} = 0\, , \\
\end{aligned}
\eeq
where
\beq
D_{z, {\zb}} {\psi} = {\pa}_{z, {\zb}} {\psi} + {A}_{z, {\zb}} {\psi}\, , \ D_{z, {\zb}} {\psi}^{\sigma} = {\pa}_{z, {\zb}} {\psi}^{\sigma} - {A}_{z, {\zb}} {\psi}^{\sigma}\, ,
\eeq
and the gauge field ${A}_{\alpha}$ and the potential $U$ are expressed through ${\psi}, {\psi}^{\sigma}$
\beq
{A}_{z, {\zb}} = - {\psi}^{\sigma} \cdot {\pa}_{z, {\zb}} {\psi}\, , \qquad
U = - \frac 12 \left( D_{z} {\psi}^{\sigma} \cdot D_{\zb} {\psi} + D_{\zb} {\psi}^{\sigma} \cdot D_{z} {\psi} \right)\, ,  
\label{eq:aau}
\eeq
which obey the usual constraint 
\beq
{\psi}^{\sigma}\cdot {\psi} = 1\ . 
\eeq
The Eqs. \eqref{eq:aau} are gauge-invariant. 
In what follows we shall often use the gauge, in which $A_{\zb} = 0$. 

\subsection{First glimpses of algebraic integrability}

{} Let us work in the real coordinates $(x,y)$ on $\Sigma$ for now. 
For the $O(N)$ model introduce the \emph{winding} ansatz:
\beq
{\psi} (x,y) =  e^{{\ii} x {\bf\theta}} f (y)  \, , \ {\psi}^{\sigma}(x,y) = f^{\sigma}(y) e^{-{\ii}x {\bf\theta}}
\label{eq:windz}
\eeq
with $f(y) \in W$, $f^{\sigma}(y) \in W^{\vee}$ for $N$ even, and, in addition,  ${\chi}(x,y) = {\chi}(y) \in {\BC}$, for $N$ odd. 
For the ${\BC\BP}^{N-1}$ model, we use \eqref{eq:windz} as well.
The twist $a$ is given by:
\beq
a = e^{\ii \bf\theta} \ .
\label{eq:atwist}
\eeq
The fields $f(y), f^{\sigma}(y)$ are constrained 
\beq
f^{\sigma} \cdot f = 1
\eeq
in the case of $O(N)$ model with even $N$ and in the case of the ${\BC\BP}^{N-1}$ model. In the $O(N)$ case
with $N$ odd
\beq
{\chi}(y)^2 + f^{\sigma}(y) \cdot f(y) = 1
\eeq
Substituting this ansatz into the equations of motion 
gives (we denote by ${\dot \Xi}$ the $y$-derivatives ${\pa}_{y}{\Xi}$)
for the $O(N)$ model for even $N$
\beq
\begin{aligned}
& {\ddot f} = \left( {\bar\tau}{\tau} {\bf\theta}^{2} - u(y) \right) f  + 2{\ii}{\tau}_{1} {\bf\theta}{\dot f} \\
& {\ddot f}^{\sigma} = f^{\sigma} \left( {\bar\tau}{\tau} {\bf\theta}^{2} - u(y) \right)   -  2{\ii}{\tau}_{1} {\dot f}^{\sigma}{\bf\theta} \\
\end{aligned}
\label{eq:neumann}
\eeq
where 
\beq
u(y) \equiv - 4{\tau}_{2}^2U =  {\ii}{\tau}_{1} \left( f^{\sigma}{\bf\theta}{\dot f} - {\dot f}^{\sigma}{\bf\theta}f \right) +
{\tau}{\bar\tau} f^{\sigma} \cdot {\bf\theta}^{2} f + {\dot f}^{\sigma} \cdot {\dot f}\ . 
\label{eq:upot}
\eeq
One can recognise in \eqref{eq:neumann},\eqref{eq:upot} a generalization of the Neumann system \cite{Neumann}, which is linearized
on the Jacobian of a spectral curve \cite{Mumford}. The system \eqref{eq:neumann} admits Lax representation with  spectral parameter, it can be mapped to a genus zero ${\mathfrak{gl}}_{2}$ Hitchin system with both regular \cite{N95} and irregular singularities. There is also an analogue of the winding ansatz for the ${\BC\BP}^{N-1}$ model, which also maps to a genus zero Hitchin system.  

There are two conclusions from this analysis. First, sigma models do have solutions described by the linear motion on some abelian variety, similar to those we saw in the quantum mechanical case. Secondly, these abelian varieties
are the Jacobians or Prym varieties of some spectral curves. 
Unfortunately, the winding ansatz does not seem to generalize in any simple way to the case of the general sigma model solutions. We need another approach.

For $N=4$ the $O(N)$ model coincides with the principal chiral field, i.e. the group $SU(2)$-valued sigma model. As any principal chiral field model, it admits the
zero curvature representation, which we discuss in \cite{KNtoap}. However, 
our approach to solving the $O(N)$ model does not use the zero-curvature type representation of equations. 
Rather, it is a development of the scheme proposed first in \cite{kr87}, then extended in \cite{kr94}, namely the construction of \emph{integrable linear operators with self-consistent potentials}.

\section{Complex Fermi-curve}\label{sec:fermi}

The construction  consists of two steps. The first step is to parameterize a periodic linear operator $-{\Delta}+u$ by {\it a spectral curve and a line bundle (divisor) on it}. The spectral curve ${\CalC}_{u}$, which we call the \emph{complex Fermi-curve}, parameterizes the complex Bloch solutions of the linear equation. The second step of the construction is the characterization of the spectral curves for which there exists a set of points on the curve such that the corresponding Bloch solutions satisfy a specific quadratic relation.

\subsection{Periodic linear operators}

In this section we present the first step for the problem in question. For a smooth double-periodic complex function $u: {\Sigma} \to {\BC}$, consider the Bloch solutions of the Schr{\"o}dinger-like equation 
\beq
{\pa}{\bar\pa} {\psi} = u (z, {\zb}) {\psi} \, ,
\label{eq:schro}
\eeq
i.e.
\beq
\begin{aligned}
& {\psi}(z+1, {\zb}+1) \ = \ a \, {\psi}(z, {\zb}) \, , \\
& {\psi}(z+{\tau}, {\zb}+{\bar\tau}) \ =\  b\,  {\psi}(z, {\zb}) \, , \\
\end{aligned}
\label{eq:blochf}
\eeq
For given $u = u(z, {\zb})$ define $C_{u} \subset {\CalM} = {\BC}^{\times}
\times {\BC}^{\times}$ to be the set of $(a,b)$ for which \eqref{eq:schro}, \eqref{eq:blochf} are obeyed. 

In \cite{kr87} it was shown that for a generic smooth periodic potential the locus  $C_{u}$ is a smooth Riemann surface of infinite genus. Moreover, it was shown that the \emph{algebraically-integrable potentials} are dense in the space of all smooth periodic potentials. The latter are the potentials for which the  normalization ${\CalC}_{u}$ of $C_{u}$ called the \emph{Fermi-curve}  is of finite genus. For such potentials ${\CalC}_{u}$ is compactified by two smooth infinity points $P_{\pm}$.
 
The Schr{\"o}dinger equation with any smooth complex potential is {\it formally self-adjoint}. Therefore to any Bloch solution with the multipliers $(a,b)\in {\CalC}_{u}$ there is the dual Bloch solution
$\psi^\sigma$ with the multipliers $(a^{-1},b^{-1})\in \CalC_u$. In other words any Fermi curve is invariant under the
holomorphic involution $\sigma: {\CalC}_{u} \to {\CalC}_{u}$:
\beq\label{eq:sigma}
{\sigma}(a,b):=(a^{-1},b^{-1})
\eeq
Note, that fixed points of the involution exist only when $E=0$ is an eigenlevel of the  (anti-) periodic problem for the operator $H$. 

\subsubsection{A model example}

It is instructive to present the simplest example with $u(z, {\zb}) = u_0= const\neq 0$. 
Let ${\Lambda}, {\bar\Lambda} \subset {\BC}$ denote the lattices, which in our world are not, in general, complex conjugate:
\beq
{\Lambda} \, = \, \{ \,  {\mt} + {\nt} {\tau} \ \vert \ {\mt},{\nt} \in {\BZ}  \} \, , \ {\bar\Lambda} \, = \, \{ \,  {\mt} + {\nt} {\bar\tau} \ \vert \ {\mt},{\nt} \in {\BZ}  \} \, , 
\qquad  {\Lambda}^{0} = {\Lambda} \backslash \{ 0 \} \,  .
\label{eq:trunclat}
\eeq
For ${\kappa} = {\mt} + {\nt} {\tau} \in {\Lambda}^{0}$,  define 
\beq
{\bar\kappa} = {\mt} + {\nt}{\bar\tau}\, , \ {\kappa}_{1} = {\mt} + {\nt} {\tau}_{1}\, , \ {\kappa}_{2} = {\nt} {\tau}_{2}\, , \label{eq:kap}
\eeq
and 
\beq
{\Lambda}^{0,\kappa} = 
{\Lambda} \backslash \{ \, 0, {\kappa} \, \}  
\eeq
The equation \eqref{eq:schro} is solved by:
\beq
{\psi}(z, {\zb},\zeta) = {\Upsilon}_{\zeta,u_0} \equiv e^{{\zeta}z + u_0 {\zeta}^{-1} {\zb}}
\label{eq:upze}
\eeq
The curve $C_{u_0}$ can be explicitly parametrized:
\beq
a({\zeta}) = e^{{\zeta} + u_0{\zeta}^{-1}} \, , \qquad
b({\zeta}) = e^{{\zeta}{\tau} + u_0{\zeta}^{-1}{\bar\tau}} \
\label{eq:sigmu0}
\eeq
It is invariant under the involution
\beq\label{invmay}
\sigma:\zeta \to -\zeta\,, \qquad a(-\zeta)=a^{-1}(\zeta)\,,\quad b(-\zeta)=b^{-1}(\zeta)
\eeq
The map \eqref{eq:sigmu0} is the normalization map of the Fermi curve $\calC_{u_0}=\BC^*$ to $C_{u_0}$. It sends an infinite number of pairs
$\left({\zeta}_{{\kappa}, u_{0}}^{-}, {\zeta}_{{\kappa}, u_{0}}^{+}\right)$, for $\kappa \in {\Lambda}^{0}$, 
to the double points $(a_{\kappa, u_{0}}, b_{{\kappa}, u_{0}}) = \left(a({\zeta}_{\kappa,u_0}^{\pm}), b({\zeta}_{\kappa,u_0}^{\pm})\right)$, where ${\zeta}_{\kappa,u_0}^{\pm}$ are the solutions of
\beq
{\zeta}_{\kappa,u_0}^{+} - {\zeta}_{\kappa,u_0}^{-} = \frac{{\pi}{\bar\kappa}}{{\tau}_{2}}\, ,  \
 {\zeta}_{\kappa,u_0}^{+} {\zeta}_{\kappa,u_0}^{-} =  \frac{\bar\kappa}{\kappa} u_0\ .
 \eeq
 Explicitly
\beq
a_{\kappa, u_0} = (-1)^{m} \, {\exp} \, \frac{{\pi}{\kappa}_{1}}{{\tau}_{2}} D_{{\kappa},u_0} \ , \quad b_{\kappa, u_0} = (-1)^{n} \, {\exp} \, \frac{{\pi} ( {\kappa}{\bar\tau} )_{1}}{{\tau}_{2}} D_{{\kappa},u_0} \, ,
\label{eq:dpab}
\eeq
where
\beq
D_{{\kappa},u_0} = \sqrt{1 + \frac{4u_0}{{\se}_{{\kappa}}}} \ ,  \quad {\zeta}_{\kappa,u_0}^{\pm} = \frac{{\pi}{\bar\kappa}}{2{\tau}_{2}} \left( {D_{{\kappa},u_0}} \pm 1 \right) \, ,
\  {\se}_{{\kappa}} = \frac{{\pi}^{2} {\kappa}{\bar\kappa}}{{\tau}{\bar\tau}}\\
\label{eq:doublep}
\eeq
and 
\beq
{\kappa}_{1} := {\mt} + {\nt} {\tau}_{1}\, ,\  ( {\kappa}{\bar\tau} )_{1} := {\mt} {\tau}_{1} + {\nt}  {\tau}{\bar\tau}\, , \
({\kappa}{\bar\tau} )_{2} : = - {\mt} {\tau}_{2}
\label{eq:kap2}
\eeq
\subsubsection{Perturbation of the curve}

Let us denote by ${\be}_{\kappa}$, for $\kappa \in \Lambda$, the double-periodic function:
\beq
{\be}_{\kappa} = {\exp} \, \frac{\pi}{\tau_2} \left( {\bar\kappa}z - {\kappa}{\zb} \right) = e^{2\pi\ii ({\mt} x + {\nt} y )}
\eeq
Now let us consider a more general potential
\beq
u = u_0 + {\ve} v = \sum_{\lambda \in \Lambda} u^{(\lambda)} {\be}_{\lambda}
\eeq
where $u^{(\lambda)} \in {\BC}$.  We view ${\ve}v = {\ve} v(z, {\zb})$ as small smooth periodic perturbation, so that
$u^{(\lambda)} = {\ve} v^{(\lambda)}$ for $\lambda \in {\Lambda}^{0}$, and $u^{(0)} = u_0 + {\ve} v^{(0)}$ for the constant mode.

We have, with ${\rm H}_{u_0} = - {\pa}{\bar\pa} + u_0$,
\beq
{\rm H}_{u_0} \left( {\be}_{\kappa} {\Upsilon}_{\zeta,u_0} \right) = E_{\kappa}({\zeta}, u_0)  \left( {\be}_{\kappa} {\Upsilon}_{\zeta,u_0} \right) \, , \eeq
 with
 \beq
E_{\kappa}({\zeta}, u_0)  = \frac{{\pi}{\kappa}}{\tau_2 \zeta} ( {\zeta} + {\zeta}_{\kappa,u_0}^{+} ) ( {\zeta} - {\zeta}_{\kappa,u_0}^{-})
\label{eq:quasien}
\eeq
The vanishing of \eqref{eq:quasien} at ${\zeta} = \mp {\zeta}^{\pm}_{\kappa,u_0}$ reflects the identity:
\beq
{\be}_{\kappa} {\Upsilon}_{{\zeta}^{-}_{\kappa,u_0},u_0} = {\Upsilon}_{{\zeta}^{+}_{\kappa,u_0},u_0}
\eeq
We expand
\beq
v = v^{(0)} + v' = \sum_{{\lambda} \in \Lambda}\ v^{(\lambda)} \, {\be}_{\lambda} \, ,
\eeq
and look for the solution of the Schr{\"o}dinger equation
\beq
\left( {\rm H}_{u_0} + {\ve} v \right) {\Psi}_{\zeta,u_0} = 0
\label{eq:pertham}
\eeq
of the form:
\beq
{\Psi}_{\zeta,u_0} = {\Upsilon}_{\zeta,u_0}  + \sum_{{\lambda} \in {\Lambda}^{0}} {\psi}^{(\lambda)}_{\zeta,u_0} \left( {\be}_{\lambda} {\Upsilon}_{\zeta,u_0}  \right)
\label{eq:psizeta}
\eeq
The equation \eqref{eq:pertham} is equivalent to the system of quadratic equations:
\begin{multline}
v^{(\kappa)} + \left( v^{(0)} + {\ve}^{-1} E_{\kappa}({\zeta},u_0) \right) {\psi}^{(\kappa)}_{\zeta,u_0}  + \sum_{{\lambda} \in {\Lambda}^{0, {\kappa}}} v^{({\lambda})} {\psi}^{({\kappa} - {\lambda})}_{\zeta,u_0}  = 0\, , \\ 
v^{(0)} + \sum_{\kappa \in {\Lambda}^{0}} v^{(\kappa)} {\psi}^{(-{\kappa})}_{\zeta, u_0} = 0 \, , \\
\label{eq:schro2}
\end{multline}
We now solve \eqref{eq:schro2} in perturbation theory, for small $\ve$. There are two types of possible scalings of the solutions:
\begin{enumerate}

\item{} Away from the double points, i.e. ${\ve} \ll | {\zeta} - {\zeta}_{\lambda,u}^{\pm} |$
 for all $\kappa \in {\Lambda}$.
In this case the solution \eqref{eq:psizeta} is dominated by the single plane-wave ${\Upsilon}_{\zeta, u_0}$, the corrections being of order $\ve$,
\beq
{\psi}^{(\lambda)}_{\zeta, u_0} = - {\ve} \frac{v^{(\lambda)}}{E_{\lambda}({\zeta},u_0)}    + \ldots, \qquad {\rm for}\ {\lambda} \in {\Lambda}^{0}\ ,
\  \label{eq:pert1wf}
\eeq
while the zero mode $u^{(0)}$ differs from $u$ by the terms of order ${\ve}^{2}$:
\beq
 u^{(0)} = u_0  - {\ve}^{2} \sum_{{\kappa} \in {\Lambda}^{0}} \frac{v^{(\kappa)}v^{(-\kappa)}}{E_{\kappa}({\zeta},u)} +\ldots \\
\label{eq:pert1en}
\eeq
with $\ldots$ denoting the higher order terms in $\ve$.
The corresponding nonsingular portion of the curve $C_{u_0}$ is deformed to $C_{u}$:
\beq
\begin{aligned}
& a({\zeta}) \ = \ {\exp}\left( {\zeta} + \frac{u^{(0)}}{\zeta} + \frac{{\tau}_{2}}{\pi} \sum_{{\kappa} \in {\Lambda}^{0}} \frac{u^{(\kappa)} u^{(-\kappa)}}{({\zeta} + {\zeta}_{\kappa,u_0}^{+}) ( {\zeta} - {\zeta}_{\kappa,u_0}^{-})}  + \ldots\right) \, \\
& b({\zeta}) \ = \
 {\exp}\left( {\tau}{\zeta}  + {\bar\tau}\frac{u^{(0)}}{\zeta} + \frac{{\tau}_{2}{\bar\tau}}{\pi} \sum_{{\kappa} \in {\Lambda}^{0}} \frac{u^{(\kappa)} u^{(-\kappa)}}{({\zeta} + {\zeta}_{\kappa,u_0}^{+}) ( {\zeta} - {\zeta}_{\kappa,u_0}^{-})} + \ldots\,\right) . \\
 \end{aligned}
 \label{eq:defcurve}
 \eeq
\item{} We zoom at the vicinity of one of the double points, i.e.
for some $\kappa \in {\Lambda}^{0}$ and a choice of "$+$" or "$-$":
${\zeta} = {\zeta}_{+}$ or ${\zeta} = {\zeta}_{-}$
where $\zeta_{\pm}$ are found from the equations
\beq
 {\zeta}_{+} - {\zeta}_{-} = \frac{{\pi}{\bar\kappa}}{\tau_2} \ ,
\ u_{-}/{\zeta}_{-} - u_{+}/{\zeta}_{+} = \frac{\pi\kappa}{\tau_2} \, , \\
\label{eq:scal1}
\eeq
or, equivalently
\beq
{\Upsilon}_{{\zeta}_{+}, u_{+}} = {\be}_{\kappa}
 {\Upsilon}_{{\zeta}_{-}, u_{-}} \ . 
 \eeq
Explicitly, $u_{\pm} = u \pm {\ve}y$, $u = u^{(0)}-{\ve}x$,
\beq
{\zeta}_{\pm} = \frac{\pi{\bar\kappa}}{2\tau_2} \left( \sqrt{D_{{\kappa}, u}^2 + \frac{4{\ve}^{2}y^{2}}{{\se}_{{\kappa}}^{2}}} - \frac{2{\ve}y}{{\se}_{{\kappa}}} \pm 1 \right)
\label{eq:scal2}
\eeq
The solution to ${\rm H}{\Psi} = 0$ can be found by expanding
\beq
{\Psi} =  {\psi}^{+} {\Upsilon}_{{\zeta}_{+}, u_{+}} + {\psi}^{-} {\Upsilon}_{{\zeta}_{-}, u_{-}} + {\ve} \sum_{{\lambda} \in {\Lambda}^{0,\kappa}} {\chi}^{\lambda} {\be}_{\lambda} {\Upsilon}_{{\zeta}_{-}, u_{-}}
\label{eq:degpert}
\eeq
where the coefficients ${\psi}^{\pm}= {\psi}^{\pm}_{0} + {\ve} {\psi}^{\pm}_{1} + \ldots , {\chi}^{\lambda} = {\chi}^{(\lambda)}_{0} + {\ve} {\chi}^{(\lambda)}_{1} + \ldots $ are to be found from the quadratic-linear equations which in the limit $\ve \to 0$ reduce to:
\beq
x^2 - y^2 = v^{(\kappa)}v^{(-\kappa)}
\label{eq:def}
\eeq
and
\begin{multline}
{\psi}^{+} \, = \, t\, v^{(\kappa)} \, =\, {\tilde t}\, (x+y) \, , \\
{\psi}^{-} \, = \, t \, (y-x) \, = \, - {\tilde t}\, v^{(-\kappa)} \, , \\
{\chi}^{(\lambda)} \, =\,  t \, \frac{(x-y) v^{(\lambda)} - v^{(\lambda - \kappa)} v^{(\kappa)}}{{\CalE}_{\lambda}({\kappa})}  \, = \\ = \, {\tilde t} \, \frac{ v^{(\lambda)}v^{(-\kappa)} - (x+y) v^{(\lambda - \kappa)}}{{\CalE}_{\lambda}({\kappa})} \, , \qquad {\lambda}  \in {\Lambda}^{0,\kappa}
\label{eq:degpert3}
\end{multline}
where $t$ or $\tilde t$ are arbitrary normalization factors,
\beq
{\CalE}_{\lambda}({\kappa}) = \frac{{\se}_{{\kappa}}}{2{\lambda}{\bar\lambda}} \left( \, 2{\lambda}{\bar\lambda}\,  - \ {\kappa}{\bar\lambda} \left( 1 +D_{{\kappa}, u^{(0)}} \right) - {\bar\kappa}{\lambda} \left( 1 -   D_{{\kappa}, u^{(0)}} \right) \right) \ .
\label{eq:clelaka}
\eeq
The portion of the curve $C_{u}$ close to the point $(a_{\kappa, u_0}, b_{\kappa, u_0})$ is parameterized by $(x,y)$ as follows, cf. \eqref{eq:kap}, \eqref{eq:kap2}:
\begin{multline}
 \frac{a(x,y)}{a_{\kappa, u^{(0)}}} \, = \, 1 + \frac{2\pi\ii\ve}{\tau_2
 {\se}_{{\kappa}}{D_{{\kappa},u^{(0)}}}} \left( \, {\kappa}_{2}\, {D_{{\kappa},u^{(0)}}}\, y - {{\kappa}_{1} }\, x \, \right)   + \ldots \, , \\
 \frac{b(x,y)}{b_{\kappa, u^{(0)}}} \,   = \, 1 + \frac{2\pi\ii\ve}{\tau_2
 {\se}_{{\kappa}}D_{{\kappa},u^{(0)}}} \left( \, ({\kappa}{\bar\tau})_{2}\, {D_{{\kappa},u^{(0)}}}\, y - ({\kappa}{\bar\tau})_{1} \, x \, \right) + \ldots
\label{eq:siguc}
\end{multline}
This is the parametrized form of a non-singular quadric, which degenerates to a pair of lines when $v^{(\kappa)}v^{(-\kappa)} \to 0$, with
the double point at $(x,y) = (0,0)$. For $v^{(\kappa)}v^{(-\kappa)} \neq 0$ the double point is resolved.
\end{enumerate}

Notice, that the resolution of double points happens simultaneously at $\kappa$ and $-\kappa$, since the parameter in the right hand side of \eqref{eq:def}
is even in $\kappa$. It is easy to see that the symmetry ${\zeta} \mapsto - \zeta$ persists at every order in perturbation theory.
\begin{rem} Note, that the perturbation theory approach used in \cite{kr87}
is of a different kind. It is necessary to emphasize that the description of the Fermi curve for finite perturbation involves resonance of higher order.
\end{rem}

\section{Algebraically integrable potentials} \label{sec:alg-int}

Recall that the algebraically-integrable potentials were defined above as those for which the Fermi curve ${\CalC}_{u}$ is of finite genus.

The theory of periodic two-dimensional operators integrable on  {\it one energy level}, goes back to the work \cite{dkn}, in which the algebraic-geometrical construction of integrable two-dimensional Schr{\"o}dinger operators in a {\it magnetic field}
\beq \label{magnetic}
H= - \frac 12 \left( D_{z} D_{\zb} + D_{\zb} D_{z} \right) + U (z,\zb).
\eeq
was proposed. The shift of the potential $U\to U-E$ transforms the equation $H\psi=E\psi$ into $H\psi=0$. Hence, without loss of generality it can be assumed that the level equals {\it zero}.

{}The construction of \cite{dkn} is based on a notion of the {\it two-point, two parametric} Baker-Akhiezer function $\psi(z,\zb,p)$. The latter is uniquely determined by a smooth genus $g$ algebraic curve $\G$ with two marked points $P_\pm$ and an effective non-special divisor  $D=\g_1+\cdots+\g_g$. The Baker-Akhiezer 
function and the operator $H$ were explicitly written in terms of the Riemann theta-function associated with the curve  $\G$.

In \cite{nv1,nv2} the {\it sufficient} conditions that single out the algebraic-geometrical data $\{\G,P_\pm,D\}$ corresponding to {\it potential} operators
\beq
H=-\pa_z\pa_\zb+ U (z,\zb)\,,
\eeq
i.e. the operators with vanishing gauge field, was found. 
The corresponding curves are the ones with a holomorphic involution $\s:\G\to \G$, having exactly {\it two} fixed points $P_\pm=\s(P_\pm)$. It is necessary to emphasize that the latter condition turns out to be crucial for another remarkable Novikov-Veselov result: the corresponding Baker-Akhiezer functions can be expressed in terms of the Prym theta-function.

In \cite{ikn} Novikov-Veselov construction was generalized for the case when the energy level is the eigenlevel of the periodic Schr{\"o}dinger operator.

Let $\Gamma$ be a smooth genus $g$ algebraic curve with $n+1$ pairs $P_{\pm}$ and $p^{(i)}_{\pm}$,  $i = 1,\ldots,n $
of punctures. We also fix the local
coordinates $k_{\pm}^{-1}(p)$ in the neighborhoods of  $P_{\pm}$, $k_{\pm}^{-1}(P_{\pm})=0$. In addition we assume a holomorphic involution $\sigma$ of the curve
\beq
\sigma: \Gamma \longmapsto \Gamma , \qquad {\sigma}\circ {\sigma} = Id\, , \label{eq:inv}
\eeq
with  ${\G}^{\sigma} = \{ P_{\pm} \} \cup \{  p^{(i)}_{\pm} \, | \,  i = 1,\ldots,n  \}$ being the set of its fixed points, i.e.
\beq
\sigma (P_{\pm})=P_{\pm}, \qquad \sigma(p_{\pm}^{(i)})= p_{\pm}^{(i)}\label{eq:fpinv}
\eeq
The local parameters are $\sigma$-odd, i.e.
\beq
k_{\pm}(\sigma (p))=-k_{\pm}(p) \label{eq:oddpar}
\eeq
The quotient ${\Gamma}/{\sigma}$ will be denoted by $\Gamma_0$. The projection
\beq
\pi :\Gamma \longmapsto \Gamma_{0}=\Gamma / \sigma \label{eq:fcurve}
\eeq
represents $\Gamma$ as a two-fold cover of $\Gamma_0$ ramified at ${\G}^{\sigma}$. In this realization the involution $\sigma$ is
a permutation of the sheets. By the Riemann-Hurwitz formula the genus of $\G$ is equal to
\beq
g=2g_0+n, \label{eq:genera}
\eeq
where $g_0$ is the genus of $\Gamma_0$.

Let $d\Omega$  be a third kind meromorphic differential on $\Gamma_0$ with poles only at
the fixed points of the involution and residues satisfying the equations
\beq\label{omegares}\Res_{P_{\pm}}d\Omega=\pm 1,\ \ \Res_{p^{(i)}_{+}}d\Omega=-\Res_{p^{(i)}_{-}}d\Omega.
\eeq
The differential $d\Omega$ has $2(g_0+n)=g+n$ zeros that will be denoted
by $\gamma_s^0,\ s=1,\ldots, g+n.$
\beq
d\Omega(\gamma_s^0)=0. \label{eq:diffdo}
\eeq
Let us choose for each $s$ a point $\gamma_s$ on $\Gamma$ such that
\beq
{\pi} ({\gamma}_{s})\, =\, {\gamma}_{s}^{0}\, , \qquad s=1,\ldots, g+n. \label{eq:preim}
\eeq
(there are $2^{g+n}$ choices). Below $\gamma_1, \ldots, \gamma_{g+n}$ will be called the admissible divisor.

\begin{lm}\cite{ikn} \label{lemBA} For generic admissible divisor $D$ there is a unique Baker-Akhiezer function  $\psi(z,\zb,p), p\in {\G}$, such that

(i) $\psi$ is meromorphic on ${\G}\setminus P_{\pm}$ and has at most simple poles at the points $\g_s$ (if they are distinct);

(ii)  in a neighborhood of the points  $P_{\pm}$ the function $\psi$ has the form
\beq\label{ii}
\psi=e^{\frac 12k_\pm(z+\zb\pm z\mp \zb)}\left( 1 + \sum_{s=1}^\infty \xi_s^{\pm}(z,\zb) k_\pm^{-s}\right), \ \ k_\pm=k_{\pm}(p);\eeq

(iii) its values at the points $p_\pm^{(i)}$ satisfy the equations
\beq\label{nod}
\psi(z,\zb,p^{(i)}_+)=\psi(z,\zb,p_-^{(i)});
\eeq

\end{lm}

\bigskip

We now recall the standard facts about the Prym variety and the Prym theta function.

\bigskip

There is a basis of $a$- and $b$-cycles on $\G$ with canonical intersection matrix: $a_i\cdot a_j=b_i\cdot b_j=0, a_i\cdot b_j=\delta_{ij}$.  In such basis the involution $\s$ acts by
\beq\label{sa}\sigma(a_i)=a_{i+g_0}, \ \ \sigma(b_i)=b_{i+g_0}, \ i=1,\ldots, g_0,
\eeq
and
\beq\label{sa1}\sigma(a_i)=-a_i, \ \ \sigma(b_i)=-b_i, \ i=2g_0+1, \ldots, 2g_0+n = g\ .
\eeq
If $d\omega_i$ are  holomorphic differentials on $\Gamma$, normalized with respect to the basis \eqref{sa},\eqref{sa1},
\beq
\oint_{a_{j}} d{\omega}_{i} = {\delta}_{i}^{j}, \qquad 1 \leq i,j \leq {g}
\eeq
then the differentials
\begin{multline} \label{prymdiff} du_i=d\omega_i-d\omega_{i+g_0},  \qquad\qquad\qquad\qquad\quad  i=1,\ldots, g_{0} \, , \\
du_i \, =\, 2 d\omega_{i+g_{0}}\, , \qquad  i=g_{0}+1, \ldots, g_{0}+n \, 
\end{multline}
are odd: ${\sigma}^*(du_{j})=-du_{j}$. By definition they are called \emph{the normalized holomorphic Prym differentials}. Note that for $n > 0$ their number $g_0 + n$ is greater than half the genus $g$ of $\G$. We denote by
\beq
d{\bu} = ( du_{j})_{j=1}^{g_0+n}
\label{eq:vecdu}
\eeq
the vector of the normalized Prym differentials.

\medskip

The matrix ${\bf\Pi} \in {\rm Mat}_{(g_{0}+n) \times (g_{0}+n)} ({\BC})$ of their $B$-periods
\beq\label{pi}
\Pi_{kj}=\oint_{b_k}du_j,\ \ 1\leq k,j\leq g_0+n\,,
\eeq
is symmetric, has positive definite imaginary part, and defines
the Prym theta-function
\beq\label{eq:prymtheta}
\theta({\bz})=\theta({\bz}|{\bf\Pi}):=\sum\limits_{{\bm}\in {\BZ}^{g_{0}+n}} e^{2\pi {\ii}({\bz},{\bm})+\pi
{\ii}(\Pi {\bm},{\bm})}\, ,
\eeq
where
for $\bz\in {\BC}^{g_{0}+n}$
\beq
 ({\bz},{\bm})=m_1z_1+\ldots+m_{g_{0}+n} z_{g_{0}+n}\ . 
 \eeq 
The theta-function has the following periodicity properties: for
\beq
{\bm}, {\bn} \in {\BZ}^{g_{0}+n} \ ,
\eeq
\beq
\label{eq:prymthper}
{\theta}( {\bz}+{\bm} + {\bf\Pi}{\bn} |{\bf\Pi}) = {\theta}({\bz}|{\bf\Pi}) e^{-2{\pi}{\ii} ({\bz} , {\bn}) - {\pi}{\ii} ( {\bn}, {\bf\Pi}{\bn} )}\ .
\eeq

\begin{lm} \label{lemma3.2}\cite{ikn} The Baker-Akhiezer function in Lemma \ref{lemBA} equals
\beq\label{eq:baf}
\psi(z,\zb,p)=\frac{ \theta({\bA}(p)+z {\bU}_{+}+\zb {\bU}_{-}+{\bZ})\,\theta({\bZ})}
{\theta(z {\bU}_{+}+\zb{\bU}_{-}+{\bZ})\, \theta({\bA}(p)+{\bZ})} \, e^{z\, {\Omega}_{+}(p)+{\zb}\, {\Omega}_{-}(p)}
\eeq
where

1) ${\bA}(p) = \int^{p}_{P_{-}} d{\bu} \in {\BC}^{g_{0}+n}/{\BZ}^{g_{0}+n} \oplus {\bf\Pi}{\BZ}^{g_{0}+n}$;

2) $\Omega_{\pm} (p)=\int^{p}_{P_{\mp}} d\Omega_\pm$ , where $d\Omega_\pm$ is a unique ${\sigma}$-odd meromorphic differential on ${\Gamma}$, normalized so that
\beq
\oint_{a_{j}} d{\Omega}_{\pm} = 0\, , \qquad j = 1, \ldots , g_{0}+n \, ,
\eeq
with the single (second order) pole at $P_{\pm}$, respectively, of the form
\beq
d\Omega_{\pm}\, = \, \left( 1+O(k_\pm^{-2})  \right)dk_{\pm} \, , \qquad p \to P_{\pm}\ .
\eeq
3) The vectors ${\bU}_{\pm} = ( U^{j}_{\pm} )_{j=1}^{g} \in {\BC}^{g}$ have the components
\beq\label{U}
U^{j}_{\pm}=\frac 1{2\pi \ii} \oint_{b_j} d\Omega_\pm \ , \qquad j = 1, \ldots , g_{0}+n\, ;
\eeq
4) the vector ${\bZ} \in {\BC}^{g_{0}+n}/{\BZ}^{g_{0}+n} \oplus {\bf\Pi}{\BZ}^{g_{0}+n}$ parametrizes the admissible divisor,
\beq
\sum_{s=1}^{g+n}{\bA}({\gamma}_{s}) +{\bZ} \in {\BZ}^{g_{0}+n} \oplus {\bf\Pi}{\BZ}^{g_{0}+n}\,,
\label{eq:divisor}
\eeq
where we recall that $d{\Omega}({\pi}({\gamma}_{s})) = 0$.
\end{lm}
\begin{rem} The definition of $\Omega_-$ above needs clarification, since $d\Omega_-$ has a pole at $P_-$. By the integral  $d\Omega_-$ we mean the choice of the  branch $\Omega_-=k_-+O(k_-^{-1})$ in a neighborhood of $P_-$ and the analytic continuation  along the path. It is assumed that the paths in the definition of $\bA(p)$ and $\Omega_\pm(p)$ are the same.
\end{rem}

\begin{thm}\cite{ikn}
The Baker-Akhiezer function $\psi(z,\zb,p)$ given by the formula \eqref{eq:baf}
satisfies the equation
\beq
(\partial_z\partial_\zb -u(z,\zb))\psi(z,\zb,p)=0, \label{eq:schr}
\eeq
with the potential
\beq
u(z,\zb)=\partial_{\zb}\xi_1^+=\partial_z \xi_1^{-}=2\partial_z\partial_\zb \ln \theta (z{\bU}^{+}+{\zb} {\bU}^{-}+{\bZ}), \label{eq:potu}
\eeq
\end{thm}
In other words, the data $ \left( {\Gamma}, {\sigma}, P_{\pm}, k_{\pm}, {\Omega} \right)$ defines the
Schr{\"o}dinger potential $u(z, {\zb})$. If it happens to be double-periodic, then its Fermi-curve and the curve $\Gamma$ coincide, ${\CalC}_{u} = {\Gamma}$.

\section{Consistency conditions}\label{sec:selfc}

\subsection{Self-consistent potentials: enters the $E$-function}

Suppose that the curve $\Gamma_0$ admits a meromorphic function $E(q)$ with $m$ simple poles. Let  us denote them by $q^{(j)}\in \Gamma_0,\ j=1,\ldots,m$, and let $\{ q^{(j)}_{+} , q^{(j)}_{-} = {\sigma}(q^{(j)}_{+}) \} = {\pi}^{-1}(q^{(j)})$ denote their preimages in $\Gamma$.  The pullback ${\pi}^{*}E$ of $E$ is the $\sigma$-even meromorphic function on ${\Gamma}$, which we also denote by $E$, so that $E \circ {\sigma} = E$. Define the ``times'' $T_{n}^{\pm}$ by expansion of $E$ near $P_{\pm}$ in the local coordinates $k_{\pm}$:
\beq\label{eq:etn}
E(p)=E_{\pm}+\sum_{n=1}^\infty T_{n}^{\pm}k_{\pm}(p)^{-2n}, \ \ p\to P_\pm \ .
\eeq
Let $\psi(z,\zb, p)$ be the Baker-Akhiezer function \eqref{eq:baf} on $\Gamma$. Define the $N=n+2m$ dimensional vector $x (z,\zb) = {\chi} \oplus {\psi} \oplus {\psi}^{\sigma}$ with ${\chi} \in {\BC}^{n}$, ${\psi} \in  {\BC}^{m}$, ${\psi}^{\sigma} \in {\BC}^{m}$ by the formulae
\beq
\begin{aligned}
&{\chi} (z,{\zb})= \left( r_{i} \, {\psi}(z,{\zb},p^{(i)}_{\pm}) \right)_{i=1}^{n}\, ,\\
& {\psi}(z, {\zb}) = \left( r_{j}\, {\psi}(z,{\zb},q^{(j)}_{+}) \right)_{j=1}^{m} \, , \\
& {\psi}^{\sigma} (z,{\zb})= \left( r_{j}\, {\psi}(z,{\zb},q^{(j)}_{-}) \right)_{j=1}^{m} \\
\end{aligned}
\eeq
with, cf. \eqref{eq:etn}
\beq
\begin{aligned}
& r_{i}^2:=-\frac{E(p_{+}^{(i)})-E(p_{-}^{(i)})}{E_{+}-E_{-}} \Res_{p_+^{(i)}} d{\Omega}\,,\ \ i=1,\ldots,n,\\
& r_j^2:=-\frac{1}{2(E_{+}-E_{-})} \Res_{q^{(j)}} Ed{\Omega}\,, \ \ j=1,\ldots,m.\\
\end{aligned}
\label{eq:normc}
\eeq
\begin{rem} The first Eq. in \eqref{eq:normc} can also be written as
\beq
r_{i}^2:=-\frac{{\Res}_{p_{+}^{(i)}} E d{\Omega} + {\Res}_{p_{-}^{(i)}} E d{\Omega} }{E_{+}-E_{-}} 
\eeq
making manifest the r{\^o}le of the differential $Ed{\Omega}$
\end{rem}
\begin{thm} The vector $x(z,\zb) \in {\BC}^{N}$ satisfies the equations
\beq\label{a}
g(x, x) \equiv \sum_{i=1}^{n} {\chi}_{i}^{2} + \sum_{j=1}^{m} {\psi}_{j} {\psi}_{j}^{\sigma}\ =\ 1 \, ,
\eeq
\beq\label{b}
g \left( {\partial}_{z} x , {\pa}_{\zb} x \right)=-u(z,\zb)
\eeq
where $u$ is the potential of the corresponding Schr{\"o}dinger operator. In addition, we have the following results for the components $T_{zz}, T_{\zb\zb}$ of the classical stress-tensor:
\beq\label{c}
\begin{aligned}
& g \left( {\pa}_{z} x , \partial_{z} x \right) \ =  T_{1}^{+}; \\
& g \left( \partial_{\zb} x , \partial_{\zb} x \right) \ = T_1^{-} \\
\end{aligned}
\eeq
as well as the expressions for the higher-spin currents
\beq\label{d}
\begin{aligned}
& g\left(\partial^2_{z} x , \partial^2_{z} x\right)\ =\, T_2^{+}+T_1^+v^+ ; \\
& g\left(\partial^2_{\zb} x , \partial^2_{\zb} x\right)\ =\,T_2^-+T_1^-v^- \\
\end{aligned}
\eeq
where
\beq
\partial_\zb v^+=u_z; \qquad \partial_z v^-=u_\zb
\eeq
\end{thm}
\noindent
{\it Proof.} Consider the differential 
\beq
d{\Omega}^{[0,0]} \, :={\psi}{\psi}^{\sigma} E d{\Omega}\, , 
\label{eq:dom00}
\eeq
where
\beq
{\psi}^{\s} (z, {\zb}, p) =  {\psi} (z,{\zb},{\sigma}(p)) \, .
\eeq
 Since the local coordinates $k_{\pm}$ in the neighborhoods of $P_{\pm}$ are $\s$-odd, the essential singularities of the first two factors in \eqref{eq:dom00} cancel. Moreover, by the definition of the admissible divisor the poles from $\psi$ and ${\psi}^{\sigma} = {\sigma}^{*}{\psi}$ are cancelled by the zeros of $d{\Omega}$.
Hence, $d{\Omega}^{[0,0]}$ is an ${\s}$-even meromorphic differential on $\G$ with the poles at where either $d\Omega$  or $E$ have poles, i.e. at  
\beq
{\G}^{\sigma} \cup \{ \, q_\pm^{(j)} \, | \, j = 1, \ldots , m \, \} \, . 
\eeq
 The sum of the residues of $d\Omega^{[0,0]}$ equals zero. Then the evaluation of the residues of $d\Omega^{[0,0]}$ proves the Eq. \eqref{a}. The Eq. \eqref{b} is a direct corollary of \eqref{a} and \eqref{eq:schr}. One can also
 prove it by the consideration of the residues of the differential $\left( {\pa}_{z}{\psi}{\pa}_{\zb}{\psi}^{\s}+
 {\pa}_{\zb}{\psi} {\pa}_{z} {\psi}^{\s} \right) Ed{\Omega}$. 

To demonstrate \eqref{c} we apply the vanishing of the sum of the residues of the meromorphic differential $d{\Omega}^{[1,1]}= {\pa}_z{\psi}{\pa}_{z}\psi^\s Ed{\Omega}$. Similarly, the Eq. \eqref{d} follows by the residue
considerations for the differential
$d{\Omega}^{[2,2]}=( {\pa}^2_{z} {\psi})({\pa}^2_{z}{\psi}^{\s}) E d{\Omega}$. $\square$

\subsection{Periodicity conditions}

In order for the potential $u(z, {\zb})$ to be double-periodic we need to impose the additional $g+n=2(g_{0}+n)$ constraints:
\beq
\begin{aligned}
& {\bU}^{+} + {\bU}^{-}  = {\bm} + {\bf\Pi}\cdot {\tilde{\bm}}, \qquad {\bm}, {\tilde\bm} \in {\BZ}^{g_{0}+n} \, \\
& {\tau}{\bU}^{+} + {\bar\tau}{\bU}^{-} = {\bl} + {\bf\Pi}\cdot {\tilde\bl}, \qquad {\bl}, {\tilde\bl} \in {\BZ}^{g_{0}+n}  \, \\
\end{aligned}
\label{eq:doubleper}
\eeq
In this case we can map the curve ${\Gamma}$ to ${\CalM} = {\BC}^{\times}\times{\BC}^{\times}$, via, cf. \eqref{eq:blochf} :
\beq
\begin{aligned}
& {\mu} : {\Gamma} \to C_{u} \subset {\CalM} \, , \ {\mu}: p \mapsto (a(p), b(p)) \\
& a(p) = e^{{\Omega}_{+}(p) + {\Omega}_{-}(p)  - 2{\pi}{\ii}
({\bA}(p), {\tilde\bm})}\, \\
& b(p) = e^{{\tau}{\Omega}_{+}(p) + {\bar\tau}{\Omega}_{-}(p)-
2{\pi}{\ii} ({\bA}(p), {\tilde\bl})} \, \\
\end{aligned}
\label{eq:blochs}
\eeq
The expression \eqref{eq:blochs} is well-defined, since the $A$-periods of $d{\Omega}_{\pm}$ are zero, the periods of $ (d{\bA}(p), {\tilde\bm})$, $(d{\bA}(p), {\tilde\bl})$ are integers, while the $B$-periods of the differentials
\beq\label{dalpha}
d\alpha:=d{\Omega}_{+}(p) + d{\Omega}_{-}(p)  - 2{\pi}{\ii} (d{\bA}(p),\tilde\bm)\, , 
\eeq
\beq\label{dbeta}
d\beta:=\ {\tau}d{\Omega}_{+}(p) + {\bar\tau}d{\Omega}_{-}(p)  - 2{\pi}{\ii} (d{\bA}(p), {\tilde\bl}) 
\eeq
are the components of the vectors $2\pi \ii{\bm}$, and $2\pi \ii{\bl}$, respectively. 

Note also that if the potential $u(z,\zb)$ is double periodic then the corresponding Baker-Akhiezer 
function is the \emph{Bloch} solution of the Schr{\"o}dinger equation, i.e.
\begin{align} \label{BABloch}
 \psi(z+1,\zb+1,p)\, =\, a(p)\psi(z,\zb,p)\,,\\  
 \psi(z+\tau,\zb+\bar \tau,p)\, =\, b(p)\psi(z,\zb,p)  \label{eq:bloch2} 
\end{align}
In order to prove \eqref{BABloch} it is enough to check that the left and right hand sides have the same analytic properties on $\G$. 

\bigskip
A smooth genus $g$ algebraic curve with involution having $2n+2$ fixed points is uniquely defined by a factor curve and a choice of $2n+2$ points of it.
Hence the space of such curves is of dimension $3g_0+2n-1$. The vectors $\bU^\pm$ above depend on a choice of the first jet of the local coordinate
$k_{\pm}^{-1}$ in the neighborhood of the marked points $P_\pm$. Hence, the total number of parameters is $3g_0+2n+1$. For fixed integer vectors $\bl,\tilde\bl,\bm,\tilde \bm$ the $2(g_{0}+n)$ Eqs. \eqref{eq:doubleper} cut out a
(local) variety of dimension $g_0+1$, if it is not empty. 

The manifold $\mathcal S^{g_0,n}$ of curves $\G$ satisfying periodicity constraints for some integer vectors $\bl,\tilde\bl,\bm,\tilde \bm$ is a union of connected components
\beq\label{SI}
{\mathcal S}^{g_0,n}=\bigcup_I \ {\mathcal S}^{g_0,n}_{I}
\eeq
The (local) coordinates on ${\mathcal S}^{g_0,n}$ can be defined similarly to those for the families of the Seiberg-Witten curves, namely 
\beq\label{Aper}
A_{0}\, = \, {\Res}_{P_+} \alpha d\beta, \qquad A_{i}\, =\, \oint_{a_i+a_{i+g_0}} \alpha d\beta\,, \quad i=1,\ldots, g_0.
\eeq
Although the Abelian integral ${\alpha}(p)$ is multi-valued the expressions \eqref{Aper} are well-defined. Indeed, a shift of $\alpha$ by a constant does not change $A_0$ since $d\beta$ has no residue. It does not change $A_i$ either, since $d\beta$ is odd with respect to the involution $\sigma$ while the cycle $a_i+a_{g+i}$ is even.

From the definition  \eqref{prymdiff}  of Prym differentials it follows, cf. \eqref{eq:blochs}:
\beq
\label{abmap}
a(p_{+}^{(i)})=a(p_{-}^{(i)})\, ,\qquad b(p_{+}^{(i)})=b(p_{-}^{(i)})
\eeq
Under the periodicity assumption the monodromy properties of the coordinates of the vector $x = {\chi} \oplus {\psi} \oplus {\psi}^{\sigma}$ are as follows:
\beq\label{moni}
\begin{aligned}
& {\chi}_{i}(z+1,\zb+1) \, = \, a_i {\chi}_{i} (z,\zb), \ \ a^2_i=a^2(p_\pm^{(i)})=1\\
& {\chi}_i(z+\tau,\zb+\bar\tau) \, =\, b_i {\chi}_i(z,\zb), \ \ b^2_i=b^2(p_\pm^{(i)})=1, 
\end{aligned}
\eeq
with $i\, =\, 1,\ldots,n$, 
while
\beq\label{monj}
\begin{aligned}
& {\psi}(z+1, {\zb}+1) = a \,{\psi}(z, {\zb})\, , \\
& {\psi}^{\sigma}(z+1, {\zb}+1) = {\psi}^{\sigma}(z, {\zb})\, a^{-1} \\
& {\psi}(z+{\tau},\zb+{\bar\tau})\ =\ b \,{\psi}(z,\zb)\, , \\
& {\psi}^{\sigma}(z+{\tau},\zb+{\bar\tau})\ =\  {\psi}^{\sigma}(z,\zb) \, b^{-1}\, , \\
\end{aligned}
\eeq
where
\beq\label{monj1}
\begin{aligned}
& a^{\pm 1} = {\rm diag} \left( a(q_{\pm}^{(1)}) , \ldots , a(q_{\pm}^{(m)}) \right) \, , \\
& b^{\pm 1} = {\rm diag} \left( b(q_{\pm}^{(1)}) , \ldots , b(q_{\pm}^{(m)}) \right) \, . \\
\end{aligned}
\eeq

\subsection{Why  $E$ exists}

The existence of the meromorphic function $E$ with certain analytical properties on the spectral curve ${\G}$ of the operator $-{\Delta} + u$ implies $u$
 is expressed quadratically in terms of its solutions. 

The goal of this section is to show that in the case of the smooth spectral curves
the existence of $E$ is also necessary, or in other words the construction above gives all the solutions of the problem in question.

Recall that the linear equation \eqref{eq:schro} follows from the variation of the Lagrangian \eqref{eq:onmodel} with respect to the variable $x$. 
The twisted boundary conditions constrain the Lagrange multiplier $U= u (z,\zb)$ by that the Fermi curve passes in the space ${\CalM}$ of Bloch multipliers, also known as the moduli space of flat ${\BC}^{\times}$-connections, through the prescribed set of points $M_{j}^{\pm} = \left( a_j^{\pm 1},b_j^{\pm 1} \right)$, determined by the twist parameters. 

In the untwisted case, $(a_{i}=b_{i}=1)$,  this becomes a highly nontrivial non-local condition that the multiplicity of the zero eigenvalue of the Schr{\"o}dinger operator equals at least $N$. 

Fix the twists 
\beq
a = \left( a_{j} \right)_{j=1}^{n+m},\, b = \left( b_{j} \right)_{j=1}^{n+m}\, , 
\eeq
with 
\beq
a_i^2 \, =\, b_i^2\, =\, 1 \, , \qquad  {\rm for}\ i=m+1,\ldots,n+m .
\eeq
let ${\CalU}_{a,b}$ be the locus of potentials $u(z,\zb)$ such that the constraint above is satisfied, i.e.
\beq\label{constab}
{\CalU}_{a,b}:=\left\{\, u \, | \ 
\left( a(q_j^\pm), b(q_j^{\pm}) \right)\ =\ ( a_j^{\pm 1},b_j^{\pm 1})\in \mathcal{C}_u\, , \ j = 1, \ldots , n+m\, \right\}
\eeq
It is stratified by the finite-dimensional loci ${\mathcal U}_{a,b}^{n,g_0}$ whose smooth open cell are the potentials constructed in the previous section, with $g_0$ the genus of the quotient-curve ${\G}_0:= {\G}/\sigma$ and $\# {\G}^{\sigma} = 2(n+1)$ fixed points of the involution. 

The locus ${\CalU}_{a,b}^{n,g_0}$ is of dimension
\beq\label{dimlocus}
\dim {\mathcal U}_{a,b}^{n,g_0}=(g_0+1-m)+(g_0+n)
\eeq
The first summand on the right hand side of \eqref{dimlocus} is the dimension
of the manifold of spectral curves in $\mathcal S^{g_0,n}$ that pass through the $2m$ non-trivial twists $a_j^{\pm 1}, b_j^{\pm}, j=1,\ldots,m$.  The second term is the dimension of the corresponding Prym variety.

The tangent space to the locus $\mathcal{U}_{a,b}$ in the infinite-dimensional space of all Schr{\"o}dinger potentials $u$ on $\Sigma$  is described by the following lemma:

\begin{lm} The variation $\delta u(z,\zb)$ is in the tangent space $T_{u} {\CalU}_{a,b}$ to the potential $u\in {\mathcal U}_{a,b}$ if and only if the equations 
\beq\label{tangentU}
\int_{\Sigma} \delta u(z,\zb)\, \psi_j(z,\zb)\psi_j^\sigma(z,\zb)dz\,d\zb=0
\eeq
hold. Here $\psi_j(z,\zb):=\psi(z,\zb,q^{(j)}_+),\  \psi_j^\sigma(z,\zb):=\psi(z,\zb,q^{(j)}_-)$.
\end{lm}
\noindent
{\it Proof.} The first variation of \eqref{eq:schro} gives the equation
\beq\label{varschro}
(\pa\bar\pa -u)\delta\psi_j=\delta u \psi_j \, . 
\eeq{}
By assumption the variation $\delta \psi_j$ has the same Bloch multipliers $(a_j,b_j)$ as $\psi_j$.
Multiplying both sides of \eqref{varschro} by the dual solution $\psi_j^\sigma$
of (\ref{eq:schro}) which has Bloch multiplies $(a_i^{-1},b_i^{-1})$,  then averaging over $\Sigma$ gives \eqref{tangentU}. 

The variation of the Lagrangian with respect to $u$ gives the equation 
\beq\label{sigmaconst}
\int_{\Sigma} \delta u(z,\zb)\, \left(1-\sum_j\psi_j\psi_j^\sigma\right)dz\,d\zb=0
\eeq
Taking into account equations \eqref{tangentU} we get that the solutions of the sigma model correspond to critical points of the functional 
\beq\label{uaver}
\left<u\right>:=\int_\Sigma u(z,\zb)dz d\zb
\eeq
restricted to the locus ${\CalU}_{a,b}$. $\square$

\begin{thm} The potential $u(z,\zb)\in {\CalU}_{a,b}$ with the smooth Fermi curve ${\CalC}_u\in \mathcal{S}^{g_0,n}$ is a critical point of the functional \eqref{uaver}
restricted onto $\mathcal U_{a,b}$ if and only if there is a meromorphic function $E$ on the quotient-curve ${\CalC}_{u,0} = {\CalC}_{u}/ {\sigma}$ with the only
simple poles at $q_j \in {\CalC}_{u,0}$, $j = 1, \ldots , m$.
\end{thm}
\noindent
{\it Proof.} First, let us show that the functional \eqref{uaver} coincides with the first coordinate $A_0$ on $\mathcal{S}^{g_0,n}$ defined in \eqref{Aper}. Namely,
\beq
\label{one}
\langle u\rangle \, = A_0 ={\Res}_{P_{+}} {\alpha} d{\beta}, 
\eeq{}
From \eqref{BABloch} it is easy to get an expression of the first coefficients of the expansions near $P_+$ of the differentials $d\alpha$ and $d\beta$
\beq\label{difexpans}
d\alpha=dk_{+}\left(\, 1+\sum_{s=1}^{\infty} {\alpha}_s\, k_{+}^{-s-1}\, \right)\,, \quad d\beta=dk_{+} \left(\, \tau+\sum_{s=1}^{\infty}\, {\beta}_{s}\, k_{+}^{-s-1}\, \right)
\eeq
in terms of the first coefficients of the expansion (\ref{ii}) of the BA function:
\beq\label{alphaxi}
\alpha_1=\xi_{1} (z,{\zb})-\xi_{1} (z+1,{\zb}+1)\,,\ 
\beta_1=\xi_{1} (z,{\zb})-\xi_{1} (z+{\tau},{\zb}+{\bar\tau})
\eeq
The Eqs. \eqref{alphaxi} imply that ${\alpha}_{1}, {\beta}_{1}$ are $(z,\zb)$-independent. From that it is easy to  get 
\beq\label{A01}
A_0=\tau \alpha_1-\beta_1=\oint_{\pa \Sigma} \xi_1 dz
=\int_\Sigma \pa_\zb \xi_1 dz\wedge d\zb
\eeq
Then using (\ref{eq:potu}) we get (\ref{one}).

\bigskip

{}Consider now the variation $\delta \alpha d\beta$. Since the periods of $d\alpha$ are constant it is a single valued meromorphic differential on the Fermi curve with at most first order poles at the marked points $P_\pm$. From \eqref{Aper} it follows that it has the form
\beq\label{deltaalph}
\delta \alpha d\beta\ =\ \delta A_0 \, \pi^*(d\Omega_{\pm}) + \sum_{i=1}^{g_0} \, \delta A_i\, \pi^*(d\omega_i)
\eeq
where $d\omega_{i}$ is the basis of normalized holomorphic differentials on ${\mathcal C}_u^{(0)} = {\mathcal C}_{u}/{\sigma}$; ${\Omega}_\pm$ is the normalized differential of the third kind with the simple poles and residues $\pm 1$ at the marked points and 
$\pi: {\mathcal C}_{u}\to {\mathcal C}_u^{(0)}$ is the projection. 

If $u(z, {\zb})$ is a critical point of the functional \eqref{uaver}, then from \eqref{one}, \eqref{deltaalph} it follows that $\delta \alpha d\beta$ is a holomorphic differential on $\mathcal C_u$. That holomorphic differential vanishes at the points $q_{j}$  for the variations tangent to the locus of spectral curves in ${\mathcal S}^{g_0,n}_{a,b}$ that are the spectral curves of the potentials $u\in {\mathcal U}^{g_0,n}_{ab}$  , since the multipliers $(a_j,b_j)$ are preserved for these deformations. In that case the dimension of the space of holomorphic differentials on ${\mathcal C}_u$ vanishing at $m$ points $q_j$ equals to the dimension of  $\mathcal S^{g_0,n}_{a,b}$ that is $g_0+1-m$. Riemann-Roch theorem gives us the dimension of the space of functions having at most simple poles at $q_j$: it is equal to $m+(g_0+1-m)-g_0+1=2$. The constants are in this space. Hence, there is a non constant function $E$ with simple poles at $q^j$
on ${\mathcal C}_u$. $\square$
\subsection{Superpotential and glimpses of Seiberg-Witten geometry}

Let us now reformulate the previous statement in the form similar to \eqref{eq:superpot}. The loci
${\CalS}^{g_{0},n}_{I}$ are the analogues of the components ${\CalU}_{\rho}$ in the quantum-mechanical case. The functional $\langle u \rangle = A_0 = {\Res}_{P_{+}} {\alpha}d{\beta}$ is the analogue of the superpotential
${\CalW}$. Moreover, by pulling the contour around $P_{+}$ so that it circles around the cuts and singularities of ${\alpha}d{\beta}$ we can make it look more like \eqref{eq:superpot}. 

\subsection{Discussion: from $O(3)$ model to reducible spectral curves} 

The construction above gives solutions of twisted $O(N)$ with  
$N=n+2m$ where $m$ is the number of poles of a meromorphic function $E$ on $\Gamma_0$. 
Since by assumption $E$ is not a constant we have $m\geq 1$. 
In this way we cover the
case of the $O(N)$ model with even $N$, and some cases of the $N$ odd. 

For fixed $m$ the solutions are indexed by $g_0$ and $I$ that is a connected component of the space  of corresponding spectral curves \eqref{SI}.

The case $N=3$ is special. There is only one possibility: $n=1$ and $m=1$. Since $m=1$ it follows $\G_0$ is a rational curve, i.e. $\G_0= {\BC\BP}^1$. Now, $\G$, being the two-sheet cover of $\G_0$ with $4$ branch points, is an elliptic curve. Therefore, the corresponding solutions  are, to some extend, trivial: they depend only on a linear combination $U z+\bar U \zb$ where $U, \bar U$ are some constants. 

{}This observation leads us \cite{KNtoap} to look for further generalizations of the Novikov-Veselov construction, namely to the case of reducible spectral curves. These, we show, produce the higher $N$ analogues of the instanton solutions
of the $O(3)$ model, in that the $T_{zz}$ and $T_{\zb\zb}$ components of the stress-tensor vanish for them. 

In \cite{KNtoap} we show that the corresponding Schr{\"o}dinger potentials satisfy the self-consistency conditions for the $O(2m+n)$-model with $n$ odd. Moreover the periodicity constraint for the potential is effectively solved in terms of the spectral curves of the elliptic Calogero-Moser system.

\subsection{Fermi curve approach to the ${\BC\BP}^{N-1}$ model}

The complex Fermi curve for the periodic two-dimensional Schr{\"o}dinger operator in a magnetic field  can be introduced in the same way \cite{dkn} in the case of the \emph{vanishing magnetic flux}, i.e. trivial principal $U(1)$-bundle $P$ or ${\BC}^{\times}$-bundle $P_{\BC}$.

\medskip
Let $\Gamma$ be a smooth genus $g$ algebraic curve with fixed local coordinates $k_\pm^{-1}$ in the neighborhoods of two marked points $P_\pm$. Let $D=\gamma_1+\cdots+\gamma_g$ be a generic effective degree $g$ divisor on $\Gamma$. Then Baker-Akhiezer function associated with this data is a unique function $\psi(z,\bar z, p)$ such that:

(i) as a function of $p\in \Gamma$ it is meromorphic on $\Gamma$ outside of the marked points $P_\pm$ with the divisor of poles $D$;

(ii) in the neighborhoods of the marked points it has the form (\ref{ii}), i.e.
$$\psi=e^{\frac 12k_\pm(z+\zb\pm z\mp \zb)}\left(\sum_{s=0}^\infty \xi_s^{\pm}(z,\zb) k_\pm^{-s}\right), \ \ k_\pm=k_{\pm}(p);
$$

(iii) normalized by the equation
\beq\label{normay}
{\xi}_0^+(z,\zb)=1
\eeq

\begin{thm} [\cite{dkn}]The BA function satisfies the equation

\beq\label{shrod28}
\left(-\partial_z\partial_{\bar z}+{A}_z(z,\bar z)\partial_{\bar z}+U(z,\bar z)\right)\psi(z,\bar z,p)=0
\eeq
where
\beq\label{Au}
{A}_z=\partial_z \ln \xi_0^-,\ \ U=\partial_{\bar z} {\xi}^+_1
\eeq
\end{thm}
\begin{rem}
The choice of the normalization \eqref{normay} of the BA function corresponds to the choice of the gauge where  ${A}_{\zb}=0$.
\end{rem}

\subsubsection{The dual BA function.} For an effective degree $g$ divisor $D$ define the dual effective degree $g$ divisor $\check D$ by the equation
\beq\label{d+d}
D+\check D={\mathcal K}+P_++P_-
\eeq
where $\mathcal K$ is the canonical class. In other words, for a generic effective degree $g$ divisor $D$ there exists a unique meromorphic differential $d\Omega$ with simple poles at $P_\pm$ and the residues  $\mp 1$ such that $d\Omega(\gamma_s)=0$. The total number of zeros of $d\Omega$ is $2g$. The points $\gamma_s^+$ are the remaining zeros of $d\Omega$. 

The dual BA function is a unique function ${\psi}^{\s}(z,\bar z,p) $ such that:

(i) as a function of $p\in \Gamma$ it is meromorphic on $\Gamma$ outside of the marked points $P_\pm$ with the divisor of poles $\check D$;

(ii) in the neighborhoods of the marked points it has the form
\beq\label{bacheck}
\psi^{\sigma} \, =\, e^{-\frac 12k_{\pm} (z+\zb\pm z\mp \zb)}\left(\sum_{s=0}^\infty {\check \xi}_s^{\pm}(z,\zb) k_\pm^{-s}\right)\, , \ k_\pm=k_{\pm}(p);
\eeq

(iii) normalized by the equation
\beq\label{normay1}
\check \xi_0^+(z,\zb)=1
\eeq

\medskip
Arguments identical to that in \cite{dkn} prove that the dual BA functions satisfies the equation
\beq\label{shrodjune}
\left(-\partial_z\partial_{\bar z}+{\check A}_z(z,\bar z)\partial_{\bar z}+ {\check U} (z,\bar z)\right) {\psi}^{\sigma} (z,\bar z,p)=0
\eeq
where
\beq\label{checkAu}
{\check{A}}_z=\partial_z \ln \check \xi_0^-,\ \ {\check U} \, =\, -\partial_{\bar z} {\check \xi}^+_1
\eeq
\noindent
From the definition of the dual BA function it follows that the differential $\psi\check \psi\, d\Omega$ is a meromorphic differential on $\Gamma$ with simple poles at $P_\pm$ with residues
\beq\label{resomega}
\Res_{P_+}\psi\check \psi d\Omega=-1,\ \ \Res_{P_-}\psi\check \psi d\Omega=\xi_0^-\, \check \xi_0^-.
\eeq
Since the sum of residues of a meromorphic differential equals zero, we get
\beq\label{c+c}
\xi_0^-=(\check \xi_0^-)^{-1}
\eeq
Similarly, we have  
\beq\label{resjune}
(\xi_1^++\check\xi_1^+)=-\Res_{P_+}(\pa_z\psi) {\psi}^{\sigma} d\Omega= \Res_{P_-}(\pa_z\psi) {\psi}^{\s}  d\Omega=(\pa_z \xi_0^-){\check \xi}_0^{-}
\eeq
Equations \eqref{shrodjune},\eqref{c+c} and \eqref{resjune} imply that the dual BA functions satisfies the formally adjoint equation
\beq\label{dualshrod28}
\left(-\partial_z\partial_{\bar z}-{A}_z(z,\bar z)\partial_{\bar z}+U(z,\bar z)-{\pa}_z {A}_z(z,\zb)\right)\check \psi(z,\bar z,p)=0
\eeq

\subsubsection{Self-consistency conditions in the ${\BC\BP}^{N-1}$ case.} Let $E(p)$ be a meromorphic function on $\Gamma$ with simple poles at the points $q_{i},\, i=1,\ldots, N$. Then $\psi \psi^{\s}\, Ed{\Omega}$ is a meromorphic differential on $\Gamma$ with the poles at the points $P_\pm$ and $q_i$. The sum of its residues is zero. Hence,

\beq\label{const28}
1=\sum_{i=1}^N r^2_i\psi_i {\psi}^{\s}_i,
\eeq
where 
\beq
\psi_i=\psi(z,\bar z,q_i)\, , \  {\psi}^{\s}_i\, =\, {\psi}^{\s}(z,\bar z,q_i)
\eeq
and
\beq
r_i^2\, =\, \frac1 {E_+-E_-}{\Res}_{q_{i}} Ed\Omega \, , \ E_\pm=E(P_\pm)
\eeq
Suppose, in addition, that the differential $dE$ vanishes at the points $P_\pm$, i.e. in the neighborhoods  of the marked points 
\beq\label{Ep}
dE(P_\pm) = 0 \quad \Rightarrow \quad  E\, =\, E_\pm+O(k_\pm^2)
\eeq{}
Then the differential $(\pa_z \psi){\psi}^{\s}\, Ed\Omega$ has no residue at the point $P_+$ while its residue
at the point $P_-$ is equal to
\beq\label{rres}
{\Res}_{P_-}(E-E_+)(\pa_z \psi)\psi^{\s} d\Omega \, =\, (E_--E_+)A_z(z,\bar z)
\eeq
Hence,
\beq\label{rres1}
\sum_i r_i^2(\pa_z\psi_i)\psi_i^{\s}\, =\, A_z
\eeq
Similarly, from the equations
\beq\label{rres2}
\Res_{P_\pm}(E-E_-)\psi\partial_{\bar z}\psi^+d\Omega=0,
\eeq
it follows that
\beq\label{rres3}
0=\sum_i r_i\psi_i \left( {\pa}_{\zb}\psi_i^{\s} \right)=-\sum_{i} r_{i} \left( {\pa}_{\zb}\psi_{i} \right)
{\psi}_i^{\s}
\eeq

\subsubsection{ Reality conditions.} Let us assume that the curve $\Gamma$ is real, i.e. there is an antiholomorphic involution $\tau:\Gamma\longmapsto \Gamma$. We will also assume that under $\tau$ the marked points are permuted, $\tau(P_\pm)=P_\mp$, and the local coordinates chosen in the neighborhood of these points satisfy the condition $k_\pm(\tau(p))=-\bar k_{\mp}(p)$.
If the divisor $D$ is real, $\tau(D)=D$, then the uniqueness of the BA function implies
\beq\label{real}
{\psi}^{\s}(p)={(\bar \xi_0^-)}^{-1}\, \bar{\psi} (\tau(p))
\eeq
From the Eqs. \eqref{real} and \eqref{c+c} it follows that 
\beq\label{cc}
\xi_0^-=\overline{ \xi_0^-}
\eeq
The differential $d\Omega$ satisfies the equation
\beq\label{realomega}
d\Omega(\tau(p))=-\overline {d\Omega}(\tau(p))
\eeq
Let as assume that
\beq\label{realE}
E(p)=-\bar E(\tau (p))
\eeq
and its poles are invariant under $\tau$, i.e, $\tau(q_i)=\tau(q_i)$. Then the constants $r_i$ in  (\ref{const28})
are real $r_i^2=\bar r_i^2$. The data can be chosen such that
$$c^2:=(\xi_0^-)^2>0, r_i^2>0$$
The gauge transformed properly normalized $\psi_i$
\beq\label{gauge}
n_i=c^{-1}r_i\psi_i
\eeq
satisfy the equations
\beq\label{shrodgauge}
(-\partial_z\partial_\zb -(\partial_\zb {\rm log}\, c)\,\partial_z+ (\partial_z {\rm log} \, c)\, \partial_{\zb}+ V)\, n_i=0
\eeq
as well as the self consistency relations
\beq\label{const28a}
\sum_{i=1}^N n_i {\bar n}_i=1,
\eeq
and, cf. \eqref{eq:aau}
\begin{multline}
A_{z} = - \sum_{i=1}^N {\bar n}_{i} (\partial_z n_i) \, = \, - \partial_z {\rm log}\, c\, , \\
A_{\zb} = - \sum_{i=1}^N {\bar n}_{i} (\partial_\zb n_i)  \, = \, \partial_\zb {\rm log}\, c
\end{multline}
i.e. they are solution of the ${\BC}{\BP}^{N-1}$ model if the periodicity constraints are satisfied. In \eqref{shrodgauge} the potential $V$ differs from the potential $U$ we discussed before by the shift by $A_z A_{\zb}$. 

\begin{rem}The explicit theta-functional formula for $\psi$ is identical to (\ref{eq:baf}) after replacing Prym theta function by Riemann theta-function defined by the matrix $\bf T$ of $b$-periods of normalized holomorphic differentials on $\Gamma$. 

The periodicity constraint for the curve $\G$ are given by the same equations
\eqref{eq:doubleper}. 
\end{rem}

\section{Conclusions and future directions} \label{sec:conc}

In this paper we found the set of complex critical points of the $O(N)$ model for even $N$ and of the ${\BC\BP}^{N-1}$ model, for generic twisted boundary conditions on the worldsheet $\Sigma$, for generic complex
metric on $\Sigma$. Each critical point extends to a semi-infinite cell in the complexification ${\CalF}^{\BC}$
of the configuration space of the corresponding sigma model, the Lefschetz thimble. The latter depends on non-holomorphic data, such as the choice of hermitian metric on ${\CalF}^{\BC}$. 

The main tool of our construction is the complex analytic curve, the Fermi-curve ${\CalC}_{u}$, which encodes the monodromy
properties of a two-dimensional Schr{\"o}dinger operator $- {\Delta} + u$. The linear fields of the sigma model are in its kernel, while their twists define a collection of points $M_{1}, \ldots, M_{N}$ in the moduli space ${\CalM}$ of flat ${\BC}^{\times}$-connections on $\Sigma$. The monodromy data (an analogue of the Riemann-Hilbert map) maps Fermi-curve to ${\CalM}$ in such a way that its image passes through $M_j$'s. 

We found that the double-periodic complex solutions of the sigma model equations of motion correspond to the linear maps of $\Sigma$ to the Prym variety of ${\CalC}_{u}$. This is a direct analogue of the result found in \cite{Nekrasov:2018pqq} for the quantum mechanical models, reviewed in the Introduction. 

We also found that the symplectic geometry of $\CalM$ is reflected in a curious way in the structure of the space of solutions. As in the quantum mechanical case \eqref{eq:superpot}, the solutions are in one-to-one correspondence with the critical points of a superpotential, which in the two dimensional case turns out to be expressed through the periods of the Seiberg-Witten-like differential ${\alpha}d{\beta}$, induced from the Atiyah-Bott symplectic form $d{\alpha} \wedge d{\beta}$ on ${\CalM}$. 

The precise connection of Fermi-curves to Seiberg-Witten curves remains a mystery. In fact, some of the sigma model solutions can be understood in terms of the analytic curves in several ways: using the twistor-like curves
corresponding to the zero curvature representation \cite{mikh-zakh, hitchin}, using Hitchin spectral curves \cite{hitchins}, and finally using our Fermi-curves. The relation between these curves is as mysterious here as it is in the conventional world of monopoles: in studying the latter on ${\BR}^{2} \times {\BS}^{1}$ one has both the 
curve which encodes the scattering data \cite{hitchinm}, and the spectral curve which actually is the Seiberg-Witten curve for quiver ${\CalN}=2$ gauge theory in four dimensions \cite{nikvas}. Despite some progress in relating them \cite{Cherkis:2007xs} the general picture is lacking.  

Several extensions and generalizations of this work are underway or should be. The case of the
$O(N)$ model with the $N$ odd leads to the reducible (singular) Fermi-curves \cite{KNtoap}. The transcedental periodicity constraints \eqref{eq:doubleper} turns out to have a remarkable representation in terms of the spectral curves of the elliptic Calogero-Moser system, which has an intimate relation both with the Seiberg-Witten theory \cite{pd,nikvas}, the theory of solitons \cite{kr-cm,trebich}, and Hitchin systems and gauge theory \cite{GN,N95}.

The construction of Lefschetz thimbles is supposed to provide the tools of evaluation of the path integral. The first step in the that direction is the analysis of the determinant of the operator of the second variation of the action. In the case of the $O(N)$ model this would be related to the determinant of our friend the Schr{\"o}dinger
operator $-{\Delta} + U$, suitably projected so that the zero modes are taken out. The complex-valuedness of the potential makes the spectrum complex, leading to the complications in the analysis of the direction of the gradient flow whose trajectories span the thimble emanating from a given critical point.

One can also envisage modifying the classical action \eqref{eq:clac} by the contribution of loops, e.g. a one-loop effective action. The critical points of the effective action, at the one-loop level, may well also be expressed
through the consistency relations of the Schr{\"o}dinger potential, relating it not only to its kernel, but also to the rest of the spectrum.   
The conventional $1/N$ analysis of the $O(N)$ and the ${\BC\BP}^{N-1}$ models done in the infinite volume \cite{DAdda:1978vbw,Witten:1978bc} predicts the breakdown of conformal invariance, mass gap generation, restoration of the global symmetry. It would be nice to verify these claims by doing a more careful analysis of the path integral. In particular, is the mass gap universal, or depends on the choice of the thimble? a linear combination thereof? Some indications of the subtlety of this problem can be found in the works \cite{Milekhin:2016fai, Gorsky1, Bolognesi:2019rwq} on the large $N$ models, although on a different geometry worldsheets $\Sigma$.

The solutions of the classical complexified sigma model on the two-torus presented in this paper are
dense in the space of all solutions in the sense that the algebraic-geometric potentials $U$, being a non-linear
generalization of trigonometric polynomials, are dense in the space of all $U$'s. We hope that the representation
of these solutions as the set of critical points of the superpotential $\CalW$ will help in evaluating the sum of the integrals over the Lefschetz thimbles. Another important goal of this project is the development of intuition about these solutions. In the quantum mechanical case the windings around the $1$-cycles in abelian variety, the complex Liouville torus, could be qualitatively understood as the gas of instantons and anti-instantons ($B$ -cycles windings ) dressed with the perturbative fluctuations ($A$-cycles windings). 
The validity of this approximation was controlled by the smallness of the parameter $e^{-{\beta}{\omega}_{0}}$, where ${\omega}_{0}$ is the frequency of the classical oscillations near the minimum of the potential, and $\beta \to \infty$ is the imaginary time. In the case of the sigma model, there is an infinite number of $\omega_0$'s, which tend to zero, thanks to the conformal invariance of the classical theory. This is the problem of point-like instantons, and it is at the origin of the breakdown \cite{bp, Polyakov:1987ez} of the instanton gas picture in the sigma models and in the four dimensional gauge theory \cite{Coleman:1978ae}. Could one find a phenomenological picture of our solutions?

\appendix

\section{A generalization of the Neumann system}

The Eqs. \eqref{eq:neumann} for even $N$ can be solved by observing that the following function of the auxiliary variable $z$ (a $U(1)$ current):
\beq
{\CalJ}(z) = \frac 12 \left( {\dot f}^{\sigma} \frac{1}{z+ {\tht}} f  - f^{\sigma} \frac{1}{z + {\tht}} {\dot f} \right) 
+ {\ii} {\tau}_{1} f^{\sigma} \frac{\tht}{z+{\tht}} f
\eeq
is conserved ${\dot {\CalJ}}(z) = 0$, for all $z$. Also, introduce
\beq
\begin{aligned}
& {\tilde A}(w) = {\dot f}^{\sigma} \frac{1}{w- {\tht}^{2}} f \, , \qquad {\tilde B}(w) = - f^{\sigma} \frac{1}{w-{\tht}^2} f \, , \\
& {\tilde C}(w) = {\dot f}^{\sigma} \frac{1}{w-{\tht}^2} {\dot f}\, , \qquad {\tilde D}(w) = - f^{\sigma} \frac{1}{w-{\tht}^2} {\dot f}
\end{aligned}
\label{eq:abcdlax}
\eeq
with $w = z^2$. 
Now, the main claim is that the spectral curve of the Lax operator 
\beq
L(w) = \left( \begin{matrix} A(w) & B(w) \\ C(w) & D(w) \end{matrix} \right)
\eeq
with $A = {\tilde A} + {\ii}{\tau}_{1} {\tilde B}_{+}$, $D = {\tilde D} + {\ii}{\tau}_{1} {\tilde B}_{+}$, $C= {\tilde C} + 2{\ii}{\tau}_{1}{\CalJ}_{+} + {\tau}{\bar\tau}$, $B = {\tilde B}$, 
where
\beq
{\tilde B}_{+} = f^{\sigma} \frac{\tht}{w-{\tht}^{2}} f\, , \qquad {\CalJ}_{+} (w) = \frac{{\CalJ}(z) + {\CalJ}(-z)}{2}
\eeq
is conserved as well:
\beq
\frac{\pa}{\pa y}  {\rm Det} \left( k - L(w) \right)  = 0
\label{eq:spectrl}
\eeq


\begin{thebibliography}{KLLSW}



\bibitem{DAdda:1978vbw}
A.~D'Adda, M.~Luscher and P.~Di Vecchia,
\emph{A $1/n$ Expandable Series of Nonlinear Sigma Models with Instantons},
Nucl. Phys. B \textbf{146}, 63-76 (1978)
doi:10.1016/0550-3213(78)90432-7


\bibitem{bp} A.~Belavin, 
A~Polyakov, \emph{Metastable States of Two-Dimensional Isotropic Ferromagnets}, 
JETP Lett. {\bf 22} (1975) 245-248


\bibitem{Bolognesi:2019rwq}
S.~Bolognesi, S.~B.~Gudnason, K.~Konishi and K.~Ohashi,
\emph{Large-$N$ ${\BC\BP}^{N-1}$ sigma model on a Euclidean torus: uniqueness and stability of the vacuum},
JHEP \textbf{12}, 044 (2019)
doi:10.1007/JHEP12(2019)044
[arXiv:1905.10555 [hep-th]].


\bibitem{Cherkis:2007xs}
S.~A.~Cherkis,
SIGMA \textbf{3}, 043 (2007)
doi:10.3842/SIGMA.2007.043
[arXiv:hep-th/0703108 [hep-th]].

\bibitem{Coleman:1978ae}
S.~R.~Coleman,
\emph{The Uses of Instantons},
Subnucl. Ser. \textbf{15}, 805 (1979)



\bibitem{Orbifolds} L.~Dixon, J.~Harvey, C.~Vafa, and E.~Witten, \emph{Strings On Orbifolds, I, II}, Nucl.~Phys.
{\bf B261} (1985) 678, B274 (1986) 285


\bibitem{dkn} 
B.A.~ Dubrovin, I.M.~ Krichever and S.P.~ Novikov, \emph{The Schr{\"o}dinger equation in a
magnetic field and Riemann surfaces}, Dokl. Akad. Nauk SSSR, 229 (1976), 15-18



\bibitem{intsys}
B.~Dubrovin, I.~Krichever, S.~Novikov, \emph{Integrable systems. I}, 
Itogi Nauki i Tekhniki, Akad. Nauk SSSR, VINITI, Dynamical Systems 4 (1985), 179-277




\bibitem{GN} A.~Gorsky, N.~Nekrasov, \emph{Elliptic Calogero-Moser system from two dimensional current algebra}, arXiv:hep-th/9401021


\bibitem{Gorsky1}  A.~Gorsky, A.~Pikalov, A.~Vainshtein,
\emph{On instability of ground states in 2D ${\BC}{\BP}^{N-1}$ and $O(N)$ models
at large $N$},  arXiv:1811.05449



\bibitem{kr-grush}
S.~Grushevsky, I.~ Krichever, \emph{Real-normalized differentials and the elliptic Calogero-Moser system},  Complex geometry and dynamics, 123–137, Abel Symp., 10, Springer, Cham, 2015.


\bibitem{Hanany:1997vm}
A.~Hanany and K.~Hori,
\emph{Branes and N=2 theories in two-dimensions},
Nucl. Phys. B \textbf{513}, 119-174 (1998)
doi:10.1016/S0550-3213(97)00754-2
[arXiv:hep-th/9707192 [hep-th]].



\bibitem{pd}
E.~D'Hoker and D.~Phong, \emph{Calogero-Moser Systems in $SU(N)$ Seiberg-Witten Theory},
Nucl.~Phys. B, 513:405-444, 1998.


\bibitem{hitchin} N.~Hitchin, \emph{Harmonic maps from 2-torus to 3-sphere}, J.~Diff.~Geom. {\bf 31} (1990) 627-710

\bibitem{hitchins} N.~Hitchin, \emph{Stable bundles and integrable systems}, Duke Math. J.
Volume {\bf 54}, Number 1 (1987), 91-114.

\bibitem{hitchinm} N.~Hitchin, \emph{Monopoles and geodesics}, Comm. Math. Phys. Volume {\bf 83}, Number 4 (1982), 579-602.

\bibitem{Hori:2000ck}
K.~Hori, A.~Iqbal and C.~Vafa,
\emph{D-branes and mirror symmetry},
[arXiv:hep-th/0005247 [hep-th]].

\bibitem{ikn} A.~Il'ina, I.~Krichever, N.~ Nekrasov, \emph{Two-dimensional periodic Schrödinger operators integrable at energy eigenlevel}, Funct. Anal. Appl. 53 (2019), no. 1, 23–36, arXiv:1903.01778v2 [math-ph]


\bibitem{mikh-zakh} V.~Mikhailov, V.~Zakharov, \emph{Relativistically invariant two-dimensional models of field
theory which are integrable by means of the inverse scattering problem method},
JETP {\bf 74} (1978) 1953-1973




\bibitem{kr77} I.~Krichever, \emph{Integration of nonlinear equations by the methods of algebraic geometry},
Funk. Anal. i Pril. 11:1 (1977), 15-31.


\bibitem{kr-cm} I.~Krichever, \emph{Elliptic solutions of the Kadomtsev-Petviashvili equation and integrable systems of particles}, Funkt. Anal. i Pril. 14 (1980), 45.



\bibitem{kr89}
I.~Krichever, \emph{Spectral theory of two-dimensional periodic operators and its applications},
Uspekhi Mat. Nauk 44:2(266) (1989), 121-184 .

\bibitem{kr87} I.~Krichever, \emph{The spectral theory of "finite-gap'' nonstationary Schrödinger operators. The nonstationary Peierls model}, Funkt. Anal. i Pril. 20:3 (1986), 42–54.

\bibitem{kr94}
I.~Krichever, {\it Algebrogeometric two-dimensional operators with self-consistent potentials}, Funct. Anal. Appl. 28 (1994), no. 1, 21–32

\bibitem{KNtoap} I.~Krichever, N.~Nekrasov, \emph{Towards Lefschets thimbles in sigma models, II}, to appear



\bibitem{Milekhin:2016fai}
A.~Milekhin,
\emph{${\BC\BP}^{N}$ sigma model on a finite interval revisited},
Phys. Rev. D \textbf{95}, no.8, 085021 (2017)
doi:10.1103/PhysRevD.95.085021
[arXiv:1612.02075 [hep-th]].

\bibitem{Mumford} D.~Mumford, \emph{Tata lectures on theta}, Birkhauser, 
Boston-Basel-Berlin 1983

\bibitem{Neumann} C.~Neumann, \emph{De problemate quodam mechanica, quod ad primam integralium ultraellipticorum classem revocatur}, Journ. f.d. reine u. angew. MAth. {\bf 56} (1859)


\bibitem{N95} N.~Nekrasov, \emph{Holomorphic bundles and many body systems}, Commun.\ Math.\ Phys.\  {\bf 180}, 587 (1996)
doi:10.1007/BF02099624 [hep-th/9503157].


\bibitem{Nekrasov:2018pqq}
N.~Nekrasov, \emph{Tying up instantons with anti-instantons}, doi:10.1142 /9789813233867$ \_$0018
arXiv:1802.04202 [hep-th].

\bibitem{nikvas} N.~Nekrasov, V.~Pestun, \emph{Seiberg-Witten geometry of four dimensional ${\CalN}=2$ quiver gauge theories},
  arXiv:1211.2240 [hep-th].


\bibitem{pohl} K.~Pohlmeyer, \emph{Integrable Hamiltonian Sysytems and Interactions through Quadratic Constraints}, Comm.~Math.~Phys. {\bf 46} (1976), 207-221


\bibitem{Polyakov:1987ez}
A.~M.~Polyakov,
\emph{Gauge Fields and Strings},
Contemp. Concepts Phys. \textbf{3}, 1-301 (1987)


\bibitem{nv1} A.P. ~Veselov, S.P. ~Novikov, \emph{Finite-zone, two-dimensional, potential Schr\"{o}dinger operators. Explicit formulas and evolution equations, } Dokl. Akad. Nauk SSSR, 279:1 (1984), 20-24;

\bibitem{nv2}
A.P. ~Veselov, S.P. ~Novikov, \emph{Finite-zone, two-dimensional Schr\"{o}dinger operators. Potential operators}, Dokl. Akad. Nauk SSSR, 279:4 (1984), 784-788;

\bibitem{trebich}  A.~Treibich,  \emph{Tangential polynomials and elliptic solitons}, Duke Math. J. {\bf 59} (3) (1989), 611\\
J. L.Verdier, \emph{New elliptic solitons}, in Algebraic Analysis 2, special volume dedicated to Prof. M.~Sato on
his 60th birthday, Academic Press, New York, 1988.


\bibitem{OrbDiscr} C.~Vafa, \emph{Modular Invariance And Discrete Torsion On Orbifolds}, Nucl. Phys. {\bf B273}
(1986) 592

\bibitem{Witten:1978bc}
E.~Witten,
\emph{Instantons, the Quark Model, and the $1/n$ Expansion},
Nucl. Phys. B \textbf{149}, 285-320 (1979)
doi:10.1016/0550-3213(79)90243-8

\bibitem{Witten:2010zr}
E.~Witten,
\emph{A New Look At The Path Integral Of Quantum Mechanics},
[arXiv:1009.6032 [hep-th]].

\bibitem{Yagi:2014toa}
J.~Yagi,
\emph{$\Omega$-deformation and quantization},
JHEP \textbf{08}, 112 (2014)
doi:10.1007/JHEP08(2014)112
[arXiv:1405.6714 [hep-th]].

\end{thebibliography}
\end{document}